\documentclass{aa}  
\usepackage{graphicx}
\usepackage{txfonts}
\usepackage{natbib}
\usepackage{dcolumn}
\newcolumntype{.}{D{.}{.}{-1}}
\newcolumntype{;}{D{;}{.}{7}}
\bibpunct{(}{)}{;}{a}{}{,} % to follow the A&A style

\newcommand{\solm}{M$_{\odot}$\ }

\begin{document}

\authorrunning{Eckart et al.}
\titlerunning{SgrA* Flares}
\title{Polarized NIR and X-ray Flares from SgrA*}
\subtitle{}
\author{A. Eckart\inst{1,2}
          \and
           F. K. Baganoff\inst{3}
          \and
          M. Zamaninasab\inst{1,2}
          \and
          M. Morris\inst{4}
          \and
          ~R. Sch\"odel\inst{1}
          \and
          ~L. Meyer\inst{1,2}
          \and
          ~K. Muzic\inst{1,2}
          \and
          M.W. Bautz\inst{3}
          \and
          W.N. Brandt\inst{5}
          \and
          G.P. Garmire\inst{5}
          \and
          ~G.R. Ricker\inst{3}
          \and
          D. Kunneriath\inst{1,2}
          \and
          ~C. Straubmeier\inst{1}
          \and
          W. Duschl\inst{6}
          \and
          ~M. Dovciak\inst{7}
          \and
          ~V. Karas\inst{7}
          \and
          ~S. Markoff\inst{8}
          \and
          ~F. Najarro\inst{9}
          \and
          ~J. Mauerhan\inst{4} 
          \and
          ~J. Moultaka\inst{10}
          \and
	  ~A. Zensus\inst{2,1}
          }
\offprints{A. Eckart (eckart@ph1.uni-koeln.de)}

  \institute{I.Physikalisches Institut, Universit\"at zu K\"oln,
             Z\"ulpicher Str.77, 50937 K\"oln, Germany\\
             \email{eckart@ph1.uni-koeln.de}
         \and
%2
             Max-Planck-Institut f\"ur Radioastronomie, 
             Auf dem H\"ugel 69, 53121 Bonn, Germany
         \and
%3
             Center for Space Research, Massachusetts Institute of
             Technology, Cambridge, MA~02139-4307, USA\\
             \email{fkb@space.mit.edu}
         \and
%4
             Department of Physics and Astronomy, University of
             California Los Angeles, Los Angeles, CA~90095-1562, USA
         \and
%5
             Department of Astronomy and Astrophysics, Pennsylvania
             State University, University Park, PA~16802-6305, USA
         \and
%6
         Institut f\"ur Theoretische Physik und Astrophysik
         Christian-Albrechts-Universit\"at zu Kiel
         Leibnizstr. 15
         24118 Kiel, Germany
         \and
%7
         Astronomical Institute, Academy of Sciences, 
         Bo\v{c}n\'{i} II, CZ-14131 Prague, Czech Republic
         \and
%8
         Astronomical Institute `Anton Pannekoek', 
         University of Amsterdam, Kruislaan 403,
         1098SJ Amsterdam, the Netherlands
         \and
%9
         Instituto de Estructura de la Materia, 
         Consejo Superior de Investigaciones Científicas, 
         CSIC, Serrano 121, 28006 Madrid, Spain
         \and
%10
         Observatoire Midi-Pyr\'en\'ees,
         14, Avenue Edouard Belin, 31400 Toulouse, France
             }

\date{Received 4 Oct. 2007 / Accepted 6 Dec. 2007}

\abstract{
Stellar dynamics indicate the presence of a super massive 
3--4$\times$10$^6$\solm ~black hole at the Galactic Center.
It is associated with the variable radio, near-infrared, and X-ray 
counterpart Sagittarius A* (SgrA*).
}{
The goal is the investigation and understanding of the physical processes 
responsible for the variable emission from SgrA*.
}{
The observations have been carried out using the NACO adaptive
optics (AO) instrument at the European Southern Observatory's Very Large
Telescope (July 2005, May 2007) and the ACIS-I instrument aboard the
\emph{Chandra X-ray Observatory} (July 2005).
}{
We find that for the July 2005 flare the variable and polarized 
NIR emission of SgrA* occurred synchronous with a moderately
bright flare event in the X-ray
domain with an excess 2 - 8 keV luminosity of about
8$\times$10$^{33}$~erg/s. 
We find no time lag between the flare events in the two wavelength bands 
with a lower limit of $\le$10~minutes.
The May 2007 flare shows the highest sub-flare to flare contrast 
observed until now.
It provides evidence for a variation in the profile of consecutive sub-flares.
}{
We confirm that highly variable and NIR polarized 
flare emission is non-thermal and that there exists
a class of synchronous NIR/X-ray flares.
We find that the flaring state can be explained via the
synchrotron self-Compton (SSC) process involving up-scattered
sub-millimeter photons from a compact source component.
The observations can be interpreted in a model involving a temporary disk 
with a short jet.
In the disk component the 
flux density variations can be explained
by spots on  relativistic orbits around the central 
super massive black hole (SMBH).
The profile variations for the May 2007 flare are interpreted as a 
variation of the spot structure due to differential rotation within the disk.
}


\maketitle
\maketitle
%
%________________________________________________________________

\section{Introduction}

At the center of our Galaxy at a distance of only $\sim$8~kpc
a super massive black hole (SMBH) of mass $\sim$4$\times$10$^6$\solm  
can convincingly be identified with the compact radio and infrared 
source Sagittarius A* (SgrA*) 
(Eckart \& Genzel 1996, Genzel et al. 1997, 2000, 
Ghez et al. 1998, 2000, 2003, 2005, 
Eckart et al. 2002, Sch\"odel et al. 2002, 2003, 
Eisenhauer 2003, 2005,
Reid \& Brunthaler 2004).
Due to its proximity SgrA* provides us with the unique opportunity 
to understand the physics and possibly the evolution of 
super massive black holes at the nuclei of galaxies.
Although, for a black hole of its size, Sgr A* is extremely under luminous
at about 10$^{-9...-10}$L$_{Edd}$
Sgr~A* is also the source of variable emission
in the X-ray and near-infrared wavelength regime
(Baganoff et al. 2001, 2003, Eckart et al.  2004, 2006a,
Porquet et al. 2003, Goldwurm et al. 2003, 
Genzel et al. 2003b, Ghez et al. 2004, Eisenhauer et al. 2005,
Belanger et al. 2006, and Yusef-Zadeh et al. 2006a).
The NIR/X-ray variability 
is probably also linked to the variability at radio
through sub-millimeter wavelengths
showing that variations occur on time scales from hours to
years (e.g. Bower et al. 2002, Herrnstein et al. 2004, Zhao et al. 2003, 2004,
Markoff, Bower \&  Falcke 2007, Markoff, Nowak \& Wilms 2005,
Mauerhan et al. 2005 and references therein).
The surprisingly low luminosity has motivated many theoretical 
and observational efforts to explain the processes that are at work 
in the immediate vicinity of Sgr A*.
For a recent summary of accretion models and variable accretion 
of stellar winds onto Sgr A* see Yuan (2006) and Cuadra \& Nayakshin (2006).

\begin{table*}
\centering
{\begin{small}
\begin{tabular}{cccll}
\hline
Telescope & Instrument & Energy/$\lambda$ & UT and JD & UT and JD \\
Observing ID & & & Start Time & Stop Time \\
\hline
VLT UT~4       & NACO      & 2.2~$\mu$m  & 2007 15 May 05:29:00 & 15 July 09:42:00  \\
               &           &             & JD 2454235.72847 & JD 2454235.90417 \\
VLT UT~4       & NACO      & 2.2~$\mu$m  & 2005 30 July 02:04:00 & 30 July 03:34:00  \\
               &           &             & JD 2453581.58611 & JD 2453581.64861 \\
\emph{Chandra} & ACIS-I    & 2-8 keV     & 2005 29 July 19:52:55 & 30 July 09:07:36  \\
               &           &             & JD 2453581.32841 & JD 2453581.88028 \\
\hline
\end{tabular}
\end{small}}
\caption{Observation log.}
\label{log}
\end{table*}

\begin{table*}
\begin{center}
\begin{small}
\begin{tabular}{llcccccccc} \hline
spectral& Date &Flare Start & Flare Stop & FWZP & FWHM & Extraction & Total & Flare & IQ state \\
domain  &      &      &      & (min)& (min)& Radius  & &  \\
& & & & & &(arcsec) & & \\ \hline
NIR    & 15 May 2007  & 7:30 $\pm$ 5 min.   & 8:30 $\pm$ 5 min. & 100$\pm$10 & $>$40      & 0.06 & 16$\pm$0.5mJy & 16$\pm$0.5mJy &  -  \\
NIR    & 30 Jul 2005  & $<$02:04            & $>$03:34          & $>$ 90     & 70$\pm$10  & 0.06 &  8$\pm$0.5mJy &  8$\pm$0.5mJy &  -  \\
X-ray  & 30 Jul 2005  & 01:59 $\pm$ 12 min. & 04:07 $\pm$ 7 min.& 128$\pm$19 & 64$\pm$10  & 1.50 &      25$\pm$3 & 18$\pm$3      & 7$\pm$1  \\
       &         &                     &                   &            &            &      &  70$\pm$8nJy  & 51$\pm$8nJy   & 20$\pm$3nJy \\
\hline
\end{tabular}
\end{small}
\end{center}
\caption{
Here we list data of the flares observed by the VLT and \emph{Chandra}.
X-ray flare data:
Given are the star and stop times and peak 
ACIS-I count rates in 10$^{-3}$ cts s$^{-1}$  and 10$^{-9}$ Jansky in 2--8~keV band of the total flare 
emission and the flare emission corrected for the count rate during
the intermediate quiescent (IQ) state. We also list the estimated start and stop times,
the full width at zero power (FWZP) and
full width at half maximum (FHWM) values, as well as the peak and IQ flux densities.
NIR flare data are given as for the X-ray flare. The peal flare flux densities
are given in 10$^{-3}$ Jansky in the NIR K$_s$-band.
}
\label{flareprop}
\end{table*}

The temporal correlation between
rapid variability of the near-infrared (NIR) and X-ray emission
(Eckart et al. 2004, Eckart et al. 2006a)
suggests that the emission showing 10$^{33-34}$~erg/s flares
arises from a compact source within a few ten
Schwarzschild radii of the SMBH.
For the SgrA* we assume 
$R_s$=2$R_g$=2GM/c$^2$$\sim$8~$\mu$as, with $R_s$ being one 
Schwarzschild radius and $R_g$ the gravitational radius of the SMBH.
By now for several simultaneous flare events
the authors found no time lag larger than an upper
limit of $\le$10~minutes, mainly given by the required binning width 
of the X-ray data.
The flaring state can be explained
with a synchrotron self-Compton (SSC) model involving up-scattered
sub-millimeter photons from a compact source component.
Inverse Compton scattering of the THz-peaked flare spectrum by the 
relativistic electrons then accounts for the X-ray emission.
This model allows for NIR flux density contributions
from both the synchrotron and SSC mechanisms.
Indications for red (Eisenhauer et al. 2005, Hornstein et al. 2006, 
Gillessen et al. 2006) 
and variable NIR spectra (Gillessen et al. 2006) 
is indicative of a possible 
exponential cutoff of the NIR/MIR synchrotron spectrum (Eckart et al. 2006a).

There is also evidence for a 17$\pm$3 minute 
quasi-periodic modulation of the NIR and X-ray emission 
(Genzel et al.2003b, Eckart et al. 2006b, Meyer et al. 2006ab, 
Belanger et al. 2006, Aschenbach et al. 2004ab).
In the following we refer to this phenomenon as QPO: 
quasi periodic oscillation.
The NIR flare emission is polarized with
a well defined range over which the position angle of the 
polarized emission is changing (60$^o$$\pm$20$^o$; Eckart et al. 2006b,
Meyer et al. 2006ab, 2007).
All these observations can be explained in a model of a temporary accretion disk harboring 
a bright orbiting spot possibly in conjunction with a short jet 
(Eckart et al. 2006b, Meyer et al. 2006ab, 2007),
suggesting a stable orientation of the source geometry over the past few years.

The millimeter/submillimeter wavelength polarization of Sgr A* 
is variable in both magnitude and position angle 
on timescales down to a few hours. 
Marrone et al. (2007)
present simultaneous observations made with the Submillimeter Array 
polarimeter at 230 and 350 GHz with sufficient sensitivity to 
determine the polarization and rotation measure at each band.
From their measurements they deduce an 
accretion rate
that does not vary by
more than 25\% and - depending on the equipartition constraints and the 
magnetic field configuration -
amounts to 2$\times$10$^{-5}$ to 2$\times$10$^{-7}$ \solm yr$^{-1}$.
The mean intrinsic position angle is 167$^\circ$$\pm$7$^\circ$
with variations of $\sim$31$^\circ$ that must originate in the 
sub-millimeter photosphere of SgrA*.

Here we present new X-ray measurements that we obtained using \emph{Chandra} and
that were taken in parallel with the NIR polarization measurements 
reported by Eckart et al. (2006b).
We also present new NIR polarization measurements taken in May 2007.
In section 2 we summarize the observations and the data reduction.
The observational results and modeling of the data 
are presented in section 3 and a more general discussion of available 
infrared and X-ray variability data on SgrA* is given in section 4.
In the Appendix we outline the used synchrotron self Compton (SSC) 
mechanism and a multi-component model for the SgrA* temporal accretion disk.
The results are summarized in section 5 and conclusions are drwan.

\section{Observations and Data Reduction}

As part of a large observing campaign Sgr~A* was observed in May 2007
and July 2005, using the VLT\footnote{Based on observations at the Very Large Telescope
(VLT) of the European Southern Observatory (ESO) on Paranal in Chile;
Programs: 073.B-0775 July 2004; 
075.B-0093 July 2005;
079.B-0084 May 2007;}.
In July 2005 we carried out simultaneously X-ray observations using
the Chandra observatory.
In the following we describe the data acquisition and reduction
for the individual telescopes.
Details of the observations are summarized in Tab.~\ref{log}.

\subsection{NIR observations and data reduction}
\label{section:NIRObservations}

The observations of SgrA* have been carried out in the NIR K$_S$-band 
(2.0-2.36$\mu$m) using 
the  NIR camera CONICA and
the adaptive optics (AO) module NAOS on the 
European Southern Observatory's Very Large
Telescope UT4
on Paranal, Chile, during the nights between 
29 and 30 July 2005 as well as 14 and 15 May 2007.
The infrared wavefront sensor of NAOS was used to lock
the AO loop on the NIR bright (K-band magnitude $\sim$6.5) supergiant
IRS~7, located about $5.6''$ north of Sgr~A*.  
Therefore the AO  was able to provide a stable correction with a high 
Strehl ratio (of the order 50\%).
In NACOS/CONICA (NACO) the combination of a Wollaston prism  
with a half-wave retarder plate 
allows the simultaneous measurement of two orthogonal 
directions of the electric field vector.

Eckart et al. (2006b) report that the variable NIR emission of SgrA* 
observed in July 2005 is 
highly polarized and consists of a contribution of a non- or weakly 
polarized main flare with highly polarized sub-flares,
showing a  possible QPO of 17$\pm$3 minutes consistent 
with previous observations.
Significant positive flux density excursions on time scales shorter 
(according to all previous NACO observations typically on the QPO time scale) 
than the duration of the overall flare 
(typically 100 minutes or more) are called sub-flares.
All further details of the observations and data reduction 
as well as an interpretation of the
July 2005 data in the framework of an orbiting spot model are given in 
Eckart et al. (2006b) and Meyer et al. (2006b).  

For the K-band polarimetry in May 2007 presented here, we took a total of  
146 frames alternating between the 0$^o$/90$^o$ and 45$^o$/135$^o$ setting.
For each image we used 4 sub-images (NDIT=4) at a 10 second 
exposure time (DIT=10s).
The images were corrected for bad pixels, sky, and flat field.
The point spread function was extracted on each individual image 
(Diolaiti et al.~\cite{diolaiti}) and then used for a Lucy-Richard 
deconvolution. After restoration with a Gaussian beam, 
aperture photometry on the diffraction limited images for 
individual sources with known flux and Sgr~A* was done.  
For the extinction correction we assumed
$A_K=2.8$\,mag. Estimates of uncertainties were obtained from the
standard deviation of fluxes from nearby constant sources. The
calibration was performed using the overall interstellar polarization
of all sources in the field, which is 4\% at $25\degr$
(Eckart et al. 1995; Ott et al. 1999).

\begin{figure}[htb]
\centering
   \includegraphics[width=7cm,angle=270]{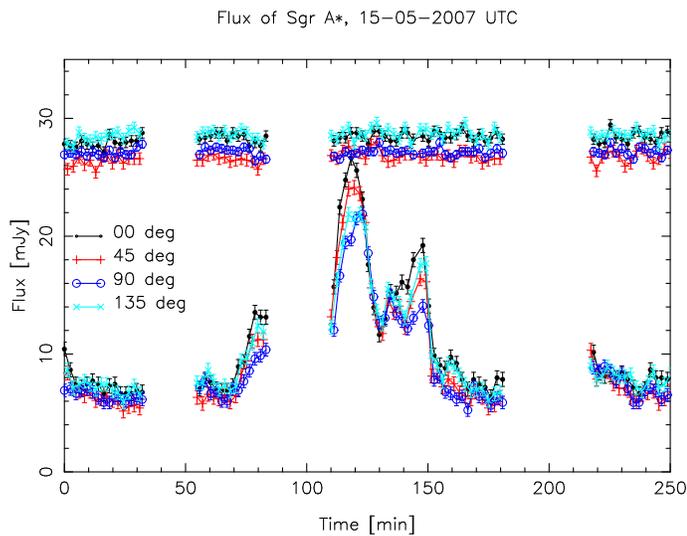}
\caption{The flux of the observed flare of Sgr A* on  15 May 2007 in
different channels as a function of time - each channel depicted 
in different color. 
Start and stop times are listed in Table~\ref{log}.
The light curve of a constant star S2 (23 mJy at 2.2 $\mu$m, with A$_K$=2.8), 
is shown in the same plot and
shifted by a few mJy for a better view. 
} \label{moha-Fig0-1}
\end{figure}

\begin{figure}[htb]
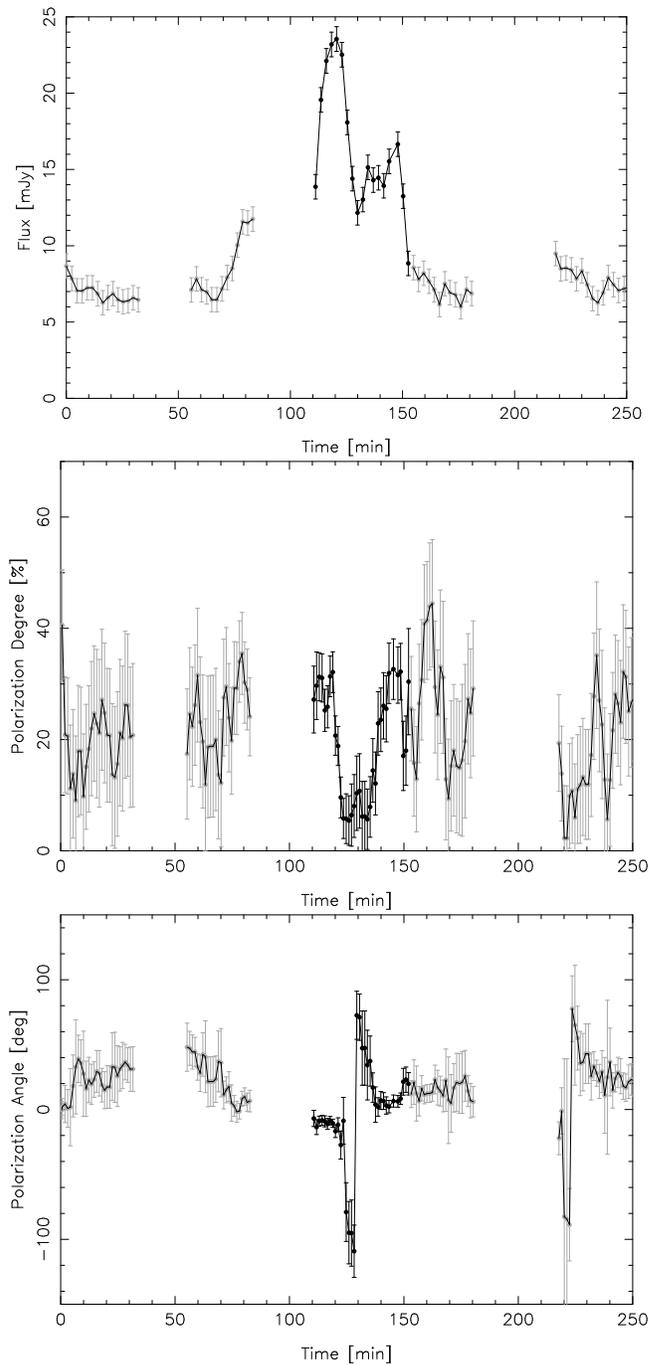

\centering
   \includegraphics[width=6cm,angle=270]{eckart8793-fig2.eps}
   \includegraphics[width=6cm,angle=270]{eckart8793-fig3.eps}
   \includegraphics[width=6cm,angle=270]{eckart8793-fig4.eps}
\caption{The SgrA* NIR flare observed on 15 May 2007.
Start and stop times are listed in Table~\ref{log}.
As a function of time we show the total flux density, 
degree of polarization and polarization angle of the E-vector.
The otherwise grey error bars are plotted in black during the time of
the largest flux excursions due to the flare.
} \label{moha-Fig0-2}
\end{figure}

In Figs.~\ref{moha-Fig0-1} and \ref{moha-Fig0-2} we show the flux densities per polarization channel,
the total intensity, degree of polarization and the polarization 
angle as a function of time.
The dereddened flux of Sgr~A* and of a nearby constant star is 
shown in the top panel of Fig.~\ref{moha-Fig0-1}. 
The flux was calibrated such that each angle separately 
matched the total flux of known sources. 
That means that actually Fig.~\ref{moha-Fig0-1} shows approximately
twice the flux for each angle. The 3 gaps in the data 
are due to sky observations. 
 
The May 2007 flare shows 2 bright sub-flares centered about 120 and 140 minutes.
The time difference between them is 20 minutes which is fully
consitent with the possible QPO found
for other polarized NIR flares of SgrA* showing a sub-flare structure.
The flux density between the 2 sub-flares does not reach the emission
level well before and after (i.e. $<$70 and $>$170 minutes into the
observations) the flare. This off-flare emission is most likely dominated
by residual stellar flux density contributions, especially from a 
faint stellar source which is currently at an apparent separation of 
 $\sim$0.1~arcsec from the position of SgrA*, but still well separated 
from it.
Based on this fact and in comparison to previously observed
flares and sub-flares
(Genzel et al.2003b, Eckart et al. 2006b, Meyer et al. 2006ab, 2007, 
Trippe et al. 2007a)
we assume that 
there
is an underlying flare emission.
The May 2007 flare shows the highest sub-flare to flare contrast 
observed until now: 
$$C=(S_{max}-S_{min})/S_{min}~~\sim 2:1~~~.$$
Here $S_{max}$ is a sub-flare flux density peak and $S_{min}$ the mean of the
neighboring minima.
The second sub-flare is broader than the first one and 
shows a flat or even double peaked profile at the top.
In section~\ref{Relativistic-disk-modeling}
we interpret these profile variations as a 
variation of the spot structure due to differential 
rotation within the disk.
Furthermore the flare is characterized by a very broad minimum in
the degree of polarization, which is preceeded by the peak of the brightest sub-flare.
During this minimum the position angle is not very well determined.
However, at the center of the minimum and 
allowing for a possible phase wrap of 180$^o$,
the trend the polarization angle shows 
towards the center of the minimum
is consistent with the value of 60$^o$$\pm$20$^o$, a value 
reported during the bright flare phase for all other flare events
(see discussion in Meyer et al. 2007).

\subsection{The \emph{Chandra} X-ray observations}
\label{section:Chandra}

In parallel to the NIR observations (Eckart et al. 2006b; see below), 
SgrA* was observed with \emph{Chandra}
using the imaging array of the Advanced
CCD Imaging Spectrometer (ACIS-I; Weisskopf et al., 2002) for
$\sim$50\,ks  on 29--30 July 2005 in the 2-8 keV band.
The start and
stop times are listed in Table~\ref{log}.
The instrument was operated in
timed exposure mode with detectors I0--3 turned on.  
The time between CCD frames was 3.141~s.  
The event data were telemetered in faint format.

We reduced and analyzed the data using CIAO v2.3\footnote{Chandra
Interactive Analysis of Observations (CIAO),
http://cxc.harvard.edu/ciao} software with Chandra CALDB
v2.22\footnote{http://cxc.harvard.edu/caldb}.  Following
Baganoff et al. (2003), we reprocessed the level~1 data to remove the
0.25\arcsec\ randomization of event positions applied during standard
pipeline processing and to retain events flagged as possible
cosmic-ray after-glows, since the strong diffuse emission in the
Galactic Center causes the algorithm to flag a significant fraction of
genuine X-rays.  The data were filtered on the standard ASCA grades.
The background was stable throughout the observation, and there were
no gaps in the telemetry.

The X-ray and optical positions of three Tycho-2 sources were
correlated (H{\o}g 2000) to register the ACIS field on the Hipparcos
coordinate frame to an accuracy of 0.10\arcsec\ (on axis); we then
measured the position of the X-ray source at Sgr~A*.  The X-ray position 
[$\alpha$$_{J2000.0}$ = $17^{\mathrm h}45^{\mathrm m}40.030^{\mathrm s}$,
$\delta$$_{J2000.0}$ = $-29^{\circ}00\arcmin28.23\arcsec$] 
is consistent with the radio position of Sgr~A* (Reid et al. 1999) to 
within $0.18\arcsec\ \pm 0.18\arcsec$ ($1\,\sigma$).

Extracting the counts from Sgr~A* in the 2--8 keV band in an 1.0\arcsec\ 
aperture provides the best compromise between maximizing
source signal and rejecting background 
(Baganoff et al. 2001, 2003, Eckart et al.  2004, 2006a).
Background counts
were extracted from an annulus around Sgr~A* with inner and outer
radii of 2\arcsec\ and 10\arcsec, respectively, excluding regions
around discrete sources and bright structures
(Baganoff et al. 2003).  
The mean (total) count rates within the inner radius subdivided into 
the peak count rates during a flare and the corresponding 
intermediate quiescent (IQ) values
are listed in Table~\ref{flareprop}.
The background rates have been
scaled to the area of the source region.  We note that the mean source
rate in the 1.5\arcsec\ aperture is consistent with the mean quiescent
source rates from previous observations 
(Baganoff et al. 2001, 2003).

 \begin{figure}
   \centering
   \includegraphics[width=8cm]{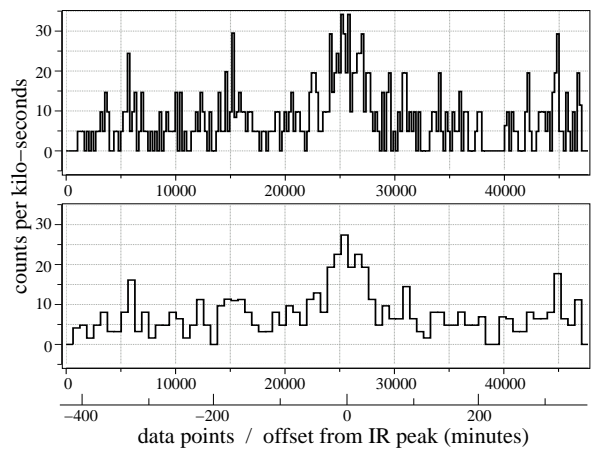}
      \caption{
%xraydata.eps
The \emph{Chandra} 2-8~keV X-ray data from 29/30 May 2005 shown in 
207 and 628 second bins.
The UT start and stop times are listed in Tab.~\ref{log}.
}
\label{fig1}
   \end{figure}

\subsection{Comparison of the NIR/X-ray flare events}
\label{section:dataresults}

\begin{figure}
   \centering
   \includegraphics[width=8cm]{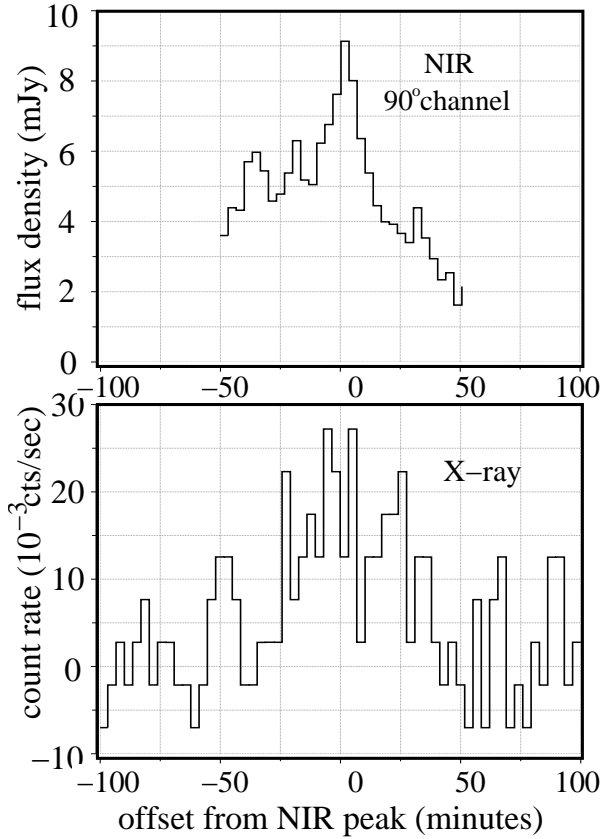}
      \caption{
%cross-data.eps;
The NIR (top) and X-ray (bottom) data for the SgrA*
flare observed on 30 May 2005. To highlight both the flare and the 
sub-flare structure we plot the flux in the NIR 90$^o$ polarization channel.
Both data sets are shown in 207 bins. 
The X-ray data are corrected for the
intermediate quiescent emission. 
The flare data are listed in Tab.~\ref{flareprop}.
The peak of the NIR fare occurred at 02:56:00 UT $\pm$3 minutes.
To within about $\pm$7 minutes the X-ray peak time occurred at the same time.
}
         \label{fig2}
   \end{figure}

\begin{figure}
   \centering
   \includegraphics[width=8cm]{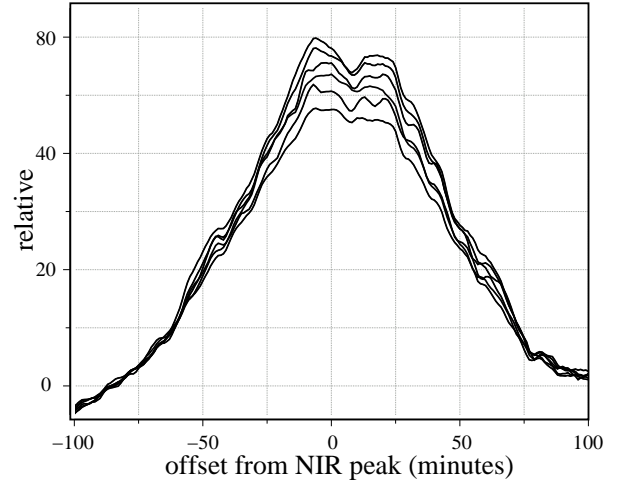}
      \caption{
%cross-alld.eps
Results of the cross correlation of the 207 second binned \emph{Chandra} X-ray data 
with the polarized emission seen at position angles of (top to bottom) 
90$^\circ$, 120$^\circ$, 60$^\circ$, 150$^\circ$, 30$^\circ$, 00$^\circ$.
}
         \label{fig3}
   \end{figure}

The \emph{Chandra} X-ray data fully cover the polarized 
NIR flare that we observed at the VLT in July 2005. 
The X-ray data show a 8$\times$10$^{33}$~erg/s flare that is about 3 times as bright
as the quiescent emission from SgrA* (Tab. \ref{flareprop}).
In Fig.\ref{fig1} 
we show the X-ray data using a 207 and a 628 second bin size. 
The cross-correlation of the X-ray data with the flux densities in the individual NIR
polarization channels show that the flare event observed in the two wavelength bands is 
simultaneous to within less than 10 minutes (Fig.\ref{fig2} and \ref{fig3}).
The two sub-peaks in the cross-correlation function correspond to two apparent 
sub-peaks in the X-ray light curve 
that can, however, not be taken as significant given the SNR of $\sim$3 cts/s per integration bin.
In the X-ray domain there is no clear indication for a QPO sub-flare structure 
as observed in the NIR.
The NIR sub-flare contrast defined as the sub-flare height divided by the height of the overall 
underlying flare flux density ranges between 0.3 and 0.9 (see section~\ref{section:SSCmodel}).
For the NIR and X-ray flare we list the start and stop times, 
flare widths, peak count rates and flux densities in Tab.~\ref{flareprop}.

\section{Modeling Results}
\label{ModelingResults}

\subsection{Relativistic disk modeling of the variable flare emission}
\label{Relativistic-disk-modeling}

We interpret our polarized infrared flare events 
via the emission of spots on relativistic 
orbits around the central SMBH in a temporary disk
(Eckart et al. 2006b, Meyer et al. 2006ab, 2007).
The model calculations are based on the KY-code by 
Dovciak, Karas, \& Yaqoob (2004)
and are usually done for a single spot 
orbiting close to the corresponding last stable orbit.
The KY-code takes special and general relativistic effects 
into account by using the concept of a transfer function
(Cunningham 1975). The transfer function relates the flux as seen by
a local observer comoving with an accretion disk to the flux 
as seen by an observer at infinity. This transfer of photons is
numerically computed by integration of the geodesic equation.
The possibility to explore effects of strong
gravity via time-resolved polarimetrical observations of X-rays was
originally proposed by Connors \& Stark (1977). Two extreme cases of the
intrinsic magnetic field configurations -- purely toroidal and purely
poloidal -- were examined as a toy model. Special and general
relativistic effects (the relativistic beaming, redshifts and
blue-shifts, lensing, time delays, the change of the emission angle and
the change of polarization angle) in the polarized light near a
Schwarzschild black hole (Pineault 1981) are calculated. 
This allows us to fit the model parameters to actual data. 
The procedure was first demonstrated by Meyer et al. (2006ab;
the authors also discuss in detail the differences to other modeling
efforts, e.g. Broderick \& Loeb 2006a,b).
The first configuration is such 
that the resulting projected E-vector is always perpendicular 
to the equatorial plane (see also Shakura \& Sunyaev 1973).   
As a second configuration we have allowed for a 
global toroidal magnetic field (Hirose et al. 2004).

Here we report on recent NIR polarimetric observations
in May 2007 which may show a sign of evolution in
the orbiting spot during the flare. 
Fitting the flare with the KY-code, we used only the total flux density and the 
degree of polarization, as the position angle of the E-vector is not 
very well determined during the minimum in the degree of polarization.

\begin{figure}[htb]
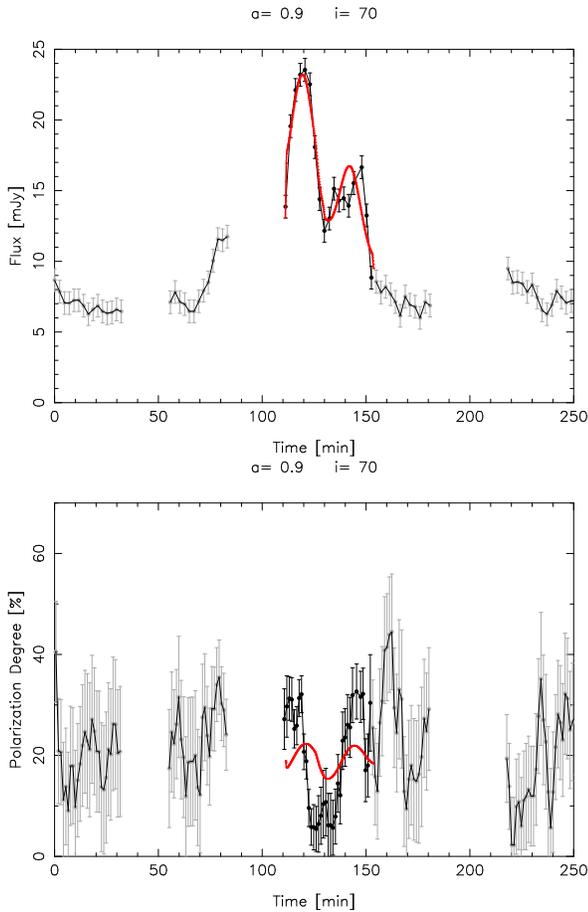

\centering
   \includegraphics[width=6cm,angle=270]{eckart8793-fig8.eps}
   \includegraphics[width=6cm,angle=270]{eckart8793-fig9.eps}
\caption{The total flux and degree of polarization for a single 
spot during two revolutions for perpendicular E-field configuration. 
The otherwise grey error bars are plotted in black during the time of
the largest flux excursions due to the flare.
The parameters of the corresponding model (I) are given in Table~\ref{Tab:KYparameters}.
} \label{moha-Fig1}
\end{figure}

Fig.~\ref{moha-Fig1} shows the fit with the least reduced 
$\chi^2$-value for synchrotron radiation, from a highly polarized spot orbiting 
twice around the SMBH above the innermost stable circular orbit (ISCO).
Since we have assumed that the spot
mainly emits synchrotron radiation the intrinsic and global
configuration of the magnetic field play an important role in the
predictions by the model. 
Here we fitted the flux and intensity of polarization
%ccccc
for the spin parameter $a$, the inclination angle $i$, an overall 
Gaussian shaped flare background, the
over-brightness of the spot according to the disk, polarization
degree  of the disk (restricted between 0\%-20\%) and the spot
(restricted between 0\%-70\%)
and the initial phase of the spot on the orbit.
The inclination $i$ is defined such that the temporary accretion disk is seen 
edge on at an inclination of $i$=90$^o$.
The dimensionless spin parameter $a$ describes the spin of the massive black hole.
A non-rotating black hole has a spin parameter $a$=0.
A maximaly spinning black hole has a spin parameter of $a$=1.
The background, the sub-flare modulation is superimposed on,
is centered at 105~minutes and has a FWHM of 80~minutes.
The upper limit for the spot 
polarization
reflects the
maximum value that could be produced by synchrotron radiation.
Fig.\ref{moha-Fig1} corresponds to a disk with a
perpendicular E-field configuration.
The least reduced-$\chi^2$ value of 4.03 is achieved for a near 
extremal SMBH ($a=0.9$) and a high inclination angle ($i=70^o$). 
The $\chi^2$ values are calculated using the data points with the 
black error bars.
Assuming that the errors are not under-estimated the fit quality,
i.e. better $\chi^2$-values, is limited by our simple physical model. 
We had to assume a factor $f_c=0.7$ by which the flux of the
components has to decrease between the first and the second revolution.
The physical origin of this factor is not clear, however, 
we assume it to be due due to 'cooling' of the spot via synchrotron losses.
The model data are plotted as a continuous smooth (red) line. 
The amplitude of the model behaves similar to observed data but 
the degree of polarization and depolarization is not so high 
during the sub-flares. 
Within this scenario it is not possible
to produce the shoulder like sub-structure of the second peak. 
Also the rise and fall in the degree of the polarization at about 120 and 140 
minutes are not sharp enough in comparison with the observations.

Within the framework of a (single) orbiting spot model, the timescale of
the signal variations, the predicted contrast of the observed light-curve
and the changes of polarization degree are constrained and mutually
interconnected. As mentioned above, by assuming that the observed
variations are dominated by the bulk orbital motion of the source, the
timescales are given by the orbital radius and the black hole angular
momentum, whereas the magnitude of the variations increases with the
source inclination and reaches its maximum for the edge-on view. It
appears that a single persisting spot is insufficient to reproduce the
observed properties of the high contrast flare event.
In agreement with the multi-component disk model presented in 
section~\ref{multicomp} 
we therefore applied another approach in which 
we employed a 2 spot model.
This double spot approach can also be interpreted in an evolutionary framework, 
motivated by the fact that accretion 
disks show magneto-rotational instabilities in which
magnetic field lines provide a coupling between disk sections at different 
radii resulting in an efficient outward transport of angular momentum.
Inner disk portions that have lost angular momentum will then slide into 
lower lying orbits, and rotate yet more rapidly
(Hawley \& Balbus 1991, Balbus 2003).
Following this idea we model the effects of differential rotation simply
by assuming two similar blobs with the same 
initial orbital phases but different radial positions 
within the extent of typical spots ($r_{spot}\sim 1-2~R_g$) that successfully 
described previous flares and sub-flares.
Also we selected these two radii such that the
mean radius is about $4.8 R_g$ according to the period of the flare
and that the sub-flare structure is fitted by the overall model. 
Such a configuration explaining the flare activity of SgrA* can 
successfully simulate a single spot that evolves in time and
separates in two entities after the first orbit.

\begin{figure}[htb]
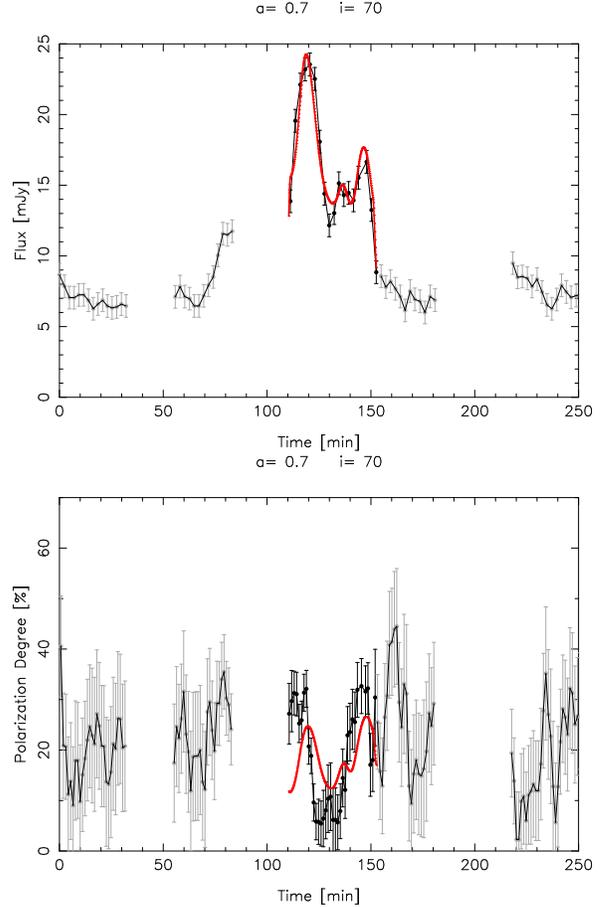

\centering
   \includegraphics[width=6cm,angle=270]{eckart8793-fig10.eps}
   \includegraphics[width=6cm,angle=270]{eckart8793-fig11.eps}
\caption{The best fit that could be achieved by a double hot spot model for
May 2007 data.
The otherwise grey error bars are plotted in black during the time of
the largest flux excursions due to the flare.
The parameters of the corresponding model (II) are given in Table~\ref{Tab:KYparameters}.
} \label{moha-Fig2}
\end{figure}

\begin{table*}
\centering
{\begin{small}
\begin{tabular}{cccccccccc}
\hline
Model  & component & flux  & radius      & spot/disk   & spin parameter & inclination \\ 
       &           &       & $r$ in r$_g$& P in \%     & $a$            & $i$         \\ 
\hline
I.     & 1         &  0.44 & 4.8         & 34/0    & 0.9        & 70$^o$         \\ 
       &           &       &             &         &            &                \\ 
II.    & 1         &  17.0 & 4.2         & 51/10   & 0.7        & 70$^o$         \\ 
       & 2         &  17.0 & 5.4         & 51/10   & 0.7        & 70$^o$         \\ 
\hline
\end{tabular}
\end{small}}
\caption{Final fit parameters resulting from the relativistic modeling 
using the KY-code.
The flux is given as the over-brightness ratio between the disk and the spot flux density
at the location of the spot in the co-moving frame. 
The polarization degree $P$ is given for the spot and disk as well.
}
\label{Tab:KYparameters}
\end{table*}

Fig.~\ref{moha-Fig2} clearly shows that this model can produce the 
substructures in flux and also the polarization degree.
The fit prefers a high inclination angle  and
spin parameter ($i=70^o$ \& $a=0.7$; reduced $\chi^2=3.55$)
which is also in good agreement
with previous observations by Eckart et al. (2006b) and
Meyer et al. (2006ab). 
We used the same assumptions for the flare background
as for the single spot model.
The two spots orbit around the
central SMBH for two revolutions with the same initial orbital phases
but different radii ($\delta r=1.2 R_g$) for the case of the 
E-field perpendicular to the disk.
As the spots separate completely from each other after the first
orbit, this model reproduces the observed sharp rises and falls 
in flux and the degree of polarization.

Finally, in the case of an azimuthal magnetic field the results 
are quite similar to the findings reported by Meyer et al. (2006ab).
We find for the best fit curves a reduced $\chi^2$-value of 4.60.
This field configuration therefore is not favored.

\section{A multi-component disk model}
\label{multicomp}

The observed NIR/X-ray properties of the SgrA* light curves raise 
a number of questions:
Can we expect a sub-flare structure in the X-ray domain
using a synchrotron self Compton model?
What is the approximate flux distribution within a temporary 
accretion disk around SgrA*?
This is also closely related to more general questions 
(discussed in section 4) of how the observed light curve properties
vary if the life time of the spot, shearing and synchrotron 
cooling time scales are considered.

In a multi-component model for the temporal accretion disk 
we combine the light amplification curves
for individual orbiting spots (based on the KY-code)
and a simple SSC model. Thereby we obtain zero order 
time dependent flare characteristics from the NIR to the 
X-ray domain.
We assume that
the essential quantities of the SSC models, i.e. the turnover flux density
$S_m$, frequency $\nu_m$, and the source size $\theta$ 
of the individual source components are distributed as power laws
with the number distributions of flux components within 
the temporal accretion disk
$$N(S) \propto S_m^{\alpha_S}~~~,~~~
N(\nu) \propto \nu_m^{\alpha_\nu}~~~,~~~
N(\theta) \propto \theta^{\alpha_\theta}~~.$$
In the Appendix
we describe the details of this extended SSC model 
that allows us to describe the disk structure (including hot spots) and to calculate 
light curves in the NIR and X-ray domain
in order to discuss the questions posed above.

We applied this multi-component disk model to the 
July 2005 and May 2007 observations presented here,
as well as the July 2004 data presented by Eckart et al. (2006a; see Tab.~\ref{Tab:SSCmulti}).

A comparison of the July 2005 X-ray data with noise sections added from the 
207~s light curve in Fig.\ref{fig1}
to the noise free panels and the light curves shown in Fig.\ref{fig2} 
demonstrates that at the given SNR and data sampling QPOs in the X-ray data
are difficult to determine, even if they have a modulation contrast
similar to that observed in the NIR.
Bright spots may on average have  smaller sizes or lower cutoff frequencies.
An increase of SSC X-ray flux density due to 
an increase of THz peak synchrotron flux may be compensated by this effect
(see expressions by Marscher et al. 1983).
Hence the sub-flare contrast may be much lower in the X-ray 
compared to the NIR domain.

\begin{figure*}[htb]
   \centering
   \includegraphics[width=16.0cm]{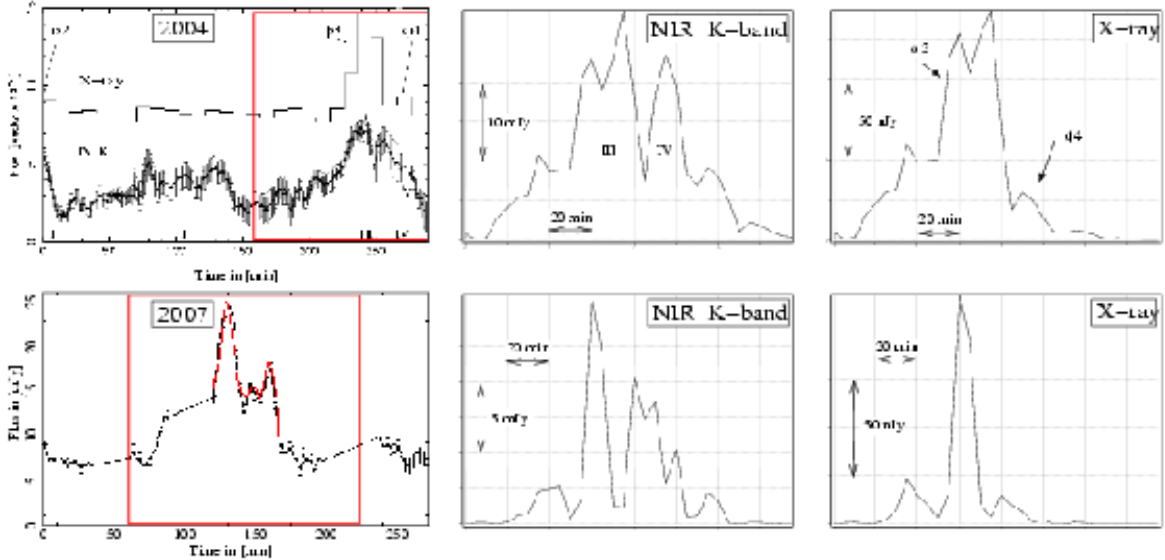}
      \caption{
%SSCdiskMODEL.eps
Application of the time dependent flare emission model presented in
section \ref{section:diskmodel} to the data obtained in 
May 2007 (bottom) and July 2004 (top).
In the panels on the left we show the available NIR and X-ray with
the modeled portion indicated by a red line. In the middle and on the
right we show the NIR K-band and X-ray light curve derived from a 
multi-component disk model. In both cases, starting at the center 
of the flare event we assumed a 30\% increase of the source component 
sizes over about 40 minutes i.e. two orbital time scales.
With the additional assumption of a flux decrease of $\sim$1 magnitude 
due to synchrotron losses the model provides a very good qualitative and
quantitative representation of the 2004 measurements (see Eckart et al. 2006a).
For the 2007 NIR data there are no simultaneous X-ray data available 
but the K-band light curve indicates an evolution of the source components.
The lower right panel therefore represents the light curve
we would have expected.
}
\label{simu20042007}
\end{figure*}

In Fig.~\ref{simu20042007} we show the modeling results for the 
May 2007 NIR and the July 2004 simultaneous NIR/X-ray data on SgrA*
using our time dependent flare emission model.
For the 2007 data we implemented a double hot spot model as described in 
section~\ref{Relativistic-disk-modeling}.
For the 2004 data we 
invoke
a model consisting of 7 components at increasing
distances from the SMBH starting at the inner last stable orbit 
(see Appendix).
The components line up on opposite sides of the SMBH close to the flare center
in time, thus providing a maximum amount of Doppler amplification 
before and afterwards.
This gives rise to the two NIR flare events labeled III and IV 
(Eckart et al. 2006a).
Motivated by the fact that the May 2007 data shows evidence for a spot evolution
due to differential rotation within the relativistic disk, we assumed that 
this may result in an increase of the source size for the individual 
spots as well.
Therefore, starting at the center of the flare event, we assumed in both cases
a 30\% increase of the source component sizes over 30 to 40 minutes,
i.e. about two orbital time scales.
This results in a sharp decrease of the SSC X-ray flux density and 
therefore in  a very good representation of the 2004 measurements 
(see X-ray flares labeled $\phi$3 and $\phi$4 in 
Fig.~\ref{simu20042007} and Eckart et al. 2006a).
Based on these time dependent model assumptions we would have expected
a similarly strong evolution of the X-ray flare light curve
for the May 2007 NIR observations as shown in the bottom right panel of
Fig.~\ref{simu20042007}. 

Such a scenario may also explain the 2006 July 17 Keck NIR/X-ray light curves
reported  by Hornstein et al. (2007).
The authors measured an NIR flare without a detectable X-ray counterpart. 
It was delayed by about 45 minutes from a significant 
X-ray flare, during which no NIR data was taken.
Assuming that the X-ray flare was accompanied by an unobserved NIR flare
as well, 
this event may have been very similar in structure to our July 2004 flare.

\section{Discussion}

While the orbiting spot model is very successful in describing 
the suggested polarized quasi periodic near-infrared flares
there are a few model assumptions that are worth being discussed 
(in section \ref{section:spotmodel})
especially with respect to a growing body of observational flare data.
To explain the feeble emission of SgrA* ($10^{-9...-11}$ of the Eddington rate) 
an intense discussion among the theoretical community at present focuses on
a combination of a radiatively inefficient accretion flow and jet models 
(e.g. Yuan, Quataert \& Narayan, 2004,
Narayan et al. 1995 
Blandford \& Begelman 1999,
Ball, Narayan, \& Quataert 2001,
Markoff et al. 2001,
Markoff, Bower \&  Falcke 2007, 
Markoff, Nowak \& Wilms 2005).
In section 
\ref{section:jetmodel}
we discuss our results in the framework of a jet model.
It is also of interest to outline links between possible
accretion disk and wind/jet models which is done in section
\ref{section:diskplusjet}.

\subsection{The orbiting spot model}
\label{section:spotmodel}

Large scale disk instabilities may be responsible for the overall
flare lengths (Tagger \& Melia 2006, Meyer et al. 2006a).
The light curves for the orbiting spot model have been calculated 
under the assumption that the spot remains confined,
i.e. to first order preserved in strength and extent,
for several orbital periods.
Here the new May 2007 data possibly represents first direct observational
evidence that spots may only be stable for about one orbital time scale.
An efficient creation of spots could be provided through
magneto-rotational instabilities that are shown to be present in Keplerian
rotating accretion disks even in the presence of a dominating
toroidal magnetic field (Hawley \& Balbus 1991, Blokland et al. 2005).
Such shear-flow instabilities are a fast mechanism to generate
a turbulent flow in a Keplerian disk (e.g. three-dimensional
simulations by Arlt \& R\"udiger 2001).
Differential rotation as well as the radial transport phenomena within the disk
stretch out the magnetic field lines that may link two disk elements.
This mechanism may provide the opportunity for field reconnection to occur.
The magnetic loops connecting the two disk elements result in poloidal fields 
and may represent the spot (Hawley \& Balbus 1991).
A spot life time that exceeds the theoretically predicted maximum of about
one orbit 
(Schnittman 2005, Schnittman et al. 2006; see discussion in section~\ref{section:variab}) 
appears to be indicated.
In the following we have assumed that in each flare the sub-flares are dominated 
by the flux density contribution of a single spot.

\subsubsection{Spot stability within the disk}
\label{section:spotstability}

If the spot is not confined it will be disintegrated via shear in the
differentially rotating disk within a very few orbital periods.
The synchrotron cooling time at 2.2$\mu$m is only a few minutes, 
i.e. significantly shorter than the time over which the spot 
may stay confined in the presence of shearing. 
If  no further heating occurs this cooling time scale will
dominate the time over which the spot persists.
The synchrotron cooling time scale $t_S$ can be calculated via 
$$t_{s, upper} < 3 \times 10^7 \Gamma \delta^{0.5} \nu^{-0.5}_9 B^{-3/2}, $$
where $t_s$ is in seconds, B is in Gauss, $\nu_9$ is frequency in GHz, 
and $\Gamma$  and $\delta$ are the relativistic factors for the 
bulk motion of the material (Blandford \& K\"onigl 1979). 
If we adopt B$\sim$ 60 G and 
$\nu_9$=300-1600~GHz 
with $\Gamma$ $\sim$ $\delta$ $\sim$ 1.5 as typical 
flare characteristics (see Eckart et al. 2006a) then
the synchrotron cooling time 
of the THz-peaked overall flare emission
is of the order of 1 to 2 hours and matches the
typical length of a flare event as observed in the K-band
as well as the 2.5 hour time scale found at 3~mm wavelength 
by Mauerhan et al. (2005).
At 2.2$\mu$m we find a much shorter cooling time scale of $T_S$$\approx$4.3~minutes.

Combining the cooling time scale with the expression for the upper
frequency of the synchrotron spectrum $\nu_2=2.8\times 10^6 B \gamma_2^2$ 
and setting it equal to the observing frequency $\nu_9$ we find that 
$$t_s \propto \nu_9^{-2}~~.$$ 
This implies that the life time of a source component bright in the 
L-band at 3.8$\mu$m will last $\approx$ 13~minutes, three times longer 
than a component seen at 2.2$\mu$m.
At the same time the upper
frequency of the synchrotron spectrum is a strong function of 
the peak synchrotron cutoff frequency and the source size. Due to 
$B \propto \theta^4 \nu_m^5 S_m^{-2}$
(e.g. Marscher 1983) we find:
$$\nu_2 \propto  B \propto  \theta^4 \nu_m^5~~.$$
Small variations in these quantities will result in a large 
variation of the upper synchrotron cutoff frequency and hence in 
a significant variation of the infrared flux density (and potentially
the infrared spectral index; see Eckart et al. 2006a).

\begin{figure}
   \centering
   \includegraphics[width=8cm]{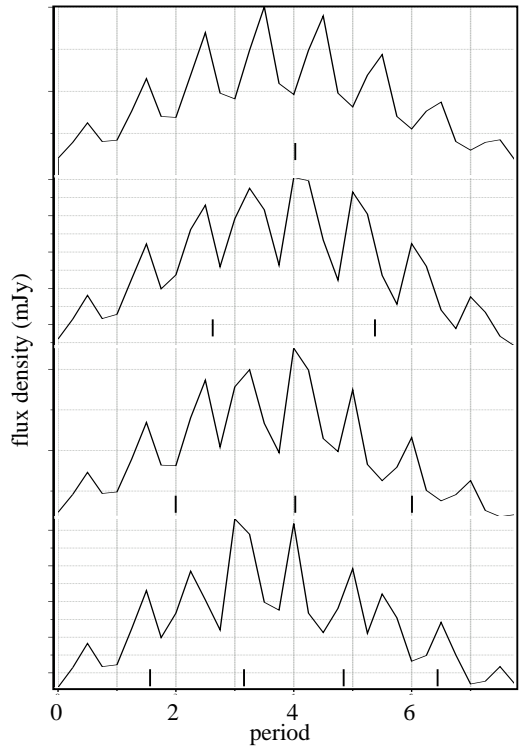}
      \caption{
%mix.eps
Simulation of 2.2$\mu$m light curves under the assumption of
decreasing times scale for the stability of source components in the
accretion disk around SgrA*.
The 
thin
vertical lines mark the centers of Gaussian shaped 
stability time intervals. These marks are spaced by a FWHM of the
individual distributions.
For short stability times scales the overall appearance of the light curve 
is preserved but the sub-flare amplitudes and time separations vary.
}
         \label{fig6}
   \end{figure}

For Fig.~\ref{fig6} 
we calculated 2.2$\mu$m light curves under the assumption of
decreasing 
times scale for the magneto-hydro-dynamical stability of source components 
in the
accretion disk around SgrA*.
The 
thin
vertical lines mark the centers of stability intervals with
Gaussian shaped flux density weights. These marks are spaced by a FWHM of the
individual Gaussians that result via modulation with the
amplification curves in light curves similar to the observed ones.
We assumed that for each of these intervals the flux density distribution 
within the disk is different. This results in phase shifts between the light curves
(i.e. different positions of the spot within the disk) of $\pm$$\pi$.
This simulation shows that the overall appearance (especially the mean QPO frequency)
of the light curve can be
preserved and that variations in the sub-flare amplitude and time separations 
can be explained by such a scenario.
In Fig.~\ref{fig6} both quantities vary by a factor of 2. Larger variations are possible
for stronger variations of the spot brightness.

In a simple model by Schnittman (2005) hot 
spots are created and destroyed around a single 
radius with random phases and exponentially 
distributed lifetimes $T_{lif}$, resulting in Lorentzian peaks 
in the power spectrum at the orbital frequency 
with a width of
$\Delta$$\nu$=(4$\pi$$T_{lif}$)$^{-1}$.
The typical lifetimes of spots in this model are 
proportional to the orbital period $T_{orb}$ at their radius. 
From MHD calculations Schnittman et al. (2006) find 
over a large range of radii that disk perturbations 
indeed have a nearly exponential distribution of lifetimes, 
with $T_{lif}$$\sim$ 0.3$T_{orb}$. 
This implies that even if the spot life time is solely determined by the
cooling time at 2.2 or 3.8~$\mu$m, this scenario is in full agreement with
the suggested quasi-periodicity since
T$_{lif}$$\sim$0.3T$_{orb}$$\sim$$T_S$ (see above
and Appendix).

From the wdith of the observed QPO of 17$\pm$3~minutes we can derive 
an expected full width of the power spectrum peak 
of $\Delta$$\nu$$\sim$0.02~min$^{-1}$. 
Following Schnittman et al. (2005) this corresponds to an
expected life time of the spots of $T_{lif}$$\sim$4~minutes
- a value similar to the synchrotron cooling time in the NIR
K-band.
However, quasi-simultaneous K- and L-band measurements by Hornstein et al. (2007) show that
for several 1 to 2 hour stretches of variable K-band emission $\ge$3~mJy, including 
flares of 10 to 30 minutes duration,
the light curves at both wavelengths are well correlated.
This suggests that the synchrotron cooling time scale in this case is not 
a relevant quantity for the spot stability.
In addition the spread $\Delta$$\nu$ is an upper limit 
to the width of a possible Lorentzian distribution describing 
the QPO measurements. 
Therefore, we have to assume that $T_{lif}$ is even longer
than the synchrotron cooling times at K- and L-band, i.e 
significantly longer than 13~minutes,
and suggesting that the spot lifetime 
could be of the order of  $T_{orb}$, in agreement with results 
by Schnittman et al. (2005) and 
the model calculations presented here.
The synchrotron cooling time scales may not be relevant at K- and L-band 
if the heating time scale is longer (e.g. on the time scale of the overall
NIR or sub-mm flare event) or if
some additional mechanism is at work that stabilizes 
the spots in the temporary accretion disk of SgrA*.
A small spot size and a high magnetic field intrinsic to the 
spot may help to prevent strong shearing, lowering the 
requirements on this confinement mechanism.

\subsection{The jet model}
\label{section:jetmodel}

Although in several cases NIR observations provide indications for QPOs 
(K-band: Genzel et al. 2003b, Eckart et al. 2006b, Meyer et al. 2006a,b
as well as in a recent L-band light curve: Trippe 2007b),
they are not apparent in all NIR light curves (e.g. see L-band light curve by 
Hornstein et al. 2007).
In addition, the radio to X-ray properties of SgrA* are often explained 
by emission from a jet
(see e.g.  Markoff, Bower \&  Falcke 2007, Markoff, Nowak \& Wilms 2005),
a source structure which is associated with 
almost all galactic nuclei.
Therefore we discuss the polarization and variability data also in the 
framework of a possible jet model.

\subsubsection{The polarization angle}
\label{section:PA}

The orientation of the polarization angle may be linked to the
intrinsic source structure.
In particular in the case of a jet
there may be a preferred orientation of the 
E-vector relative to the jet orientation.
Pollack, Taylor \& Zayala (2003) find for a sample of 177 sources a flat distribution 
of position angles with a tendency of the E-vector being perpendicular to the jet direction.
Rusk (1988) and Gabuzda \& Cawthorne (2000) find a weak indication for parallel orientations
in stronger beamed jet sources.
The mean 230 and 345~GHz
intrinsic position angle is 167$^\circ$$\pm$7$^\circ$
(E of N)
with variations of $\sim$31$^\circ$ (Marrone et al. 2007). 
Within the error this position angle is orthogonal
to that of the NIR polarized emission at about 60-80$^\circ$ (E of N). 
Both position angles also show a similar amount of variability 
(about 30$^\circ$; Eckart et al. 2006b, Meyer et al. 2006a,b).
Alternating orthogonal polarization angles are a common feature observed 
in jets.
It is also very likely that the flare spectrum of SgrA* is THz-peaked. 
In this case orthogonal polarization angles between frequencies above (NIR) and below
(mm/sub-mm domain) the synchrotron cutoff frequency are expected.
However, the model by Liu et al. (2007) also
explains the millimeter and sub-millimeter polarization properties 
with the emission originating entirely from a hot accretion disk.
This supports that the emission is from small regions and therefore associated 
with flare events occurring either in coronae of the disk or within the last stable orbit. 
It also shows that the polarization angle itself is not a sufficient indicator
to decide between a jet and disk structure.

\subsubsection{Variability and synchrotron cooling}
\label{section:variab}

\begin{figure}
   \centering
   \includegraphics[width=8cm]{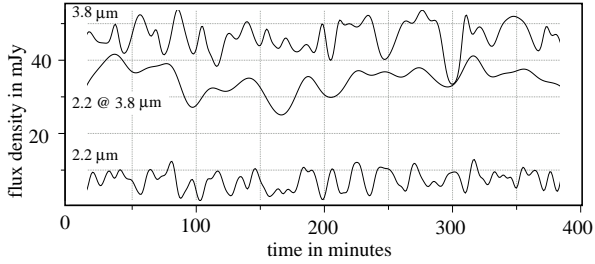}
      \caption{
%jetfig.eps
Simulated light curves at 2.2$\mu$m and 3.8$\mu$m wavelength.
The time scale has been set such that the rate of positive flux density
excursions matches approximately the variability observed at 2.2$\mu$m. 
The smooth central line shows the contribution of the 2.2$\mu$m events
at 3.8$\mu$m. All additional flux density variations seen at 3.8$\mu$m
are due to lower energy components radiating longward of 2.2$\mu$m.   
}
         \label{fig7}
   \end{figure}

If the emission is originates in the foot point of a jet,
we would expect that the time dependency 
of the flux density is not necessarily inferred via the amplification curves of the orbiting spot 
model but solely from jet instabilities and the synchrotron cooling. 
Here we investigate the effect of the the synchrotron cooling and show in Fig.\ref{fig7} 
simulated simultaneous light curves at 2.2$\mu$m and 3.8$\mu$m.
For a mean magnetic field of B$\sim$60~G we took the frequency dependence of the
synchrotron cooling time into account. The light curve was calculated using the
SSC formalism described in the Appendix
resulting in source components with different 
upper synchrotron cutoff frequencies. We also assumed that the synchrotron 
heating time is shorter than the cooling time and that at any time the 
NIR/MIR spectrum is dominated by a single source component. 
Source components that are bright at
$\le$2.2$\mu$m are bright at 3.8$\mu$m  for a correspondingly longer time.
For source components that only appear at 3.8$\mu$m we assumed a minimum cooling time
of 3 minutes.

The middle curve in Fig.\ref{fig7} (labeled 2.2$\mu$m@3.8$\mu$m) shows the 
3.8$\mu$m flux density contribution of the components that are bright at 2.2$\mu$m 
and shorter.
All additional flux density contributions that are included in the 3.8$\mu$m
light curve are due to source components with synchrotron cutoffs at wavelengths 
longward of 2.4$\mu$m.
Fig.\ref{fig7} demonstrates that, if the synchrotron cooling time is the only
effect that is responsible for the observed flux density variations, then the light curves 
at longer infrared wavelengths will be strongly influenced by all lower energy 
synchrotron events and will not at all resemble the light curves observed 
at shorter NIR wavelengths.
The quasi-simultaneous K- and L-band measurements by Hornstein et al. (2007) show that
the K- and L-band light curves at both wavelengths are well correlated,
suggesting that the synchrotron cooling time scale in this case is not relevant.
This implies that the heating time scale was longer or the component responsible for the 
flare event was stabilized by some mechanism (see section~\ref{section:spotstability}).

\subsubsection{The K-band flare rate}

\begin{figure}
   \centering
   \includegraphics[width=7cm]{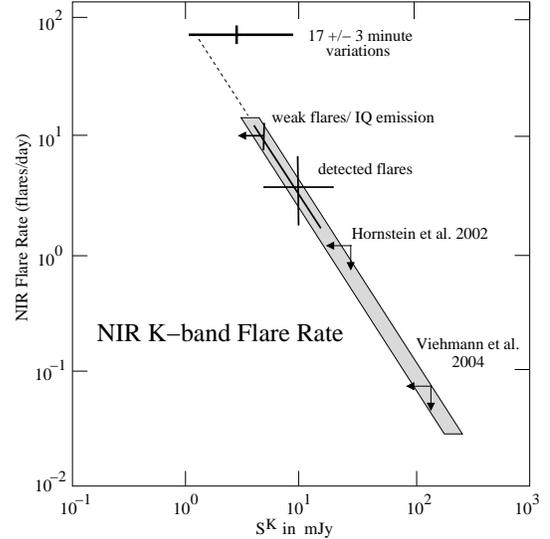}
      \caption{
%flareplot2007.eps
Flare rate as a function of flare amplitude
for the NIR K-band emission 
from SgrA* under the assumption that the characteristic flare duration is 
of the order of 100 minutes (see Eckart et al. 2006a). The 17$\pm$3 minute
variations observed as the sub-flare structure lies close to the 
extrapolation of the power law line derived from flare measurements.
It includes all NIR flare events (see references in 
section~\ref{section:jetmodel})
as well as a K-band event with high sub-flare contrast observed in
May 2007.
}
         \label{fig8}
   \end{figure}

In case of a jet, the 17$\pm$3 minute infrared 
(optically thin) flux density variations may more 
likely be a result of the variations in the accretion process 
(or jet instabilities) rather than 
being a result of a modulation from an orbiting spot. 
In this case one may expect that red noise 
variations on these short times scales are a natural extension 
of the variability found for longer periods.
Fig.~\ref{fig8} shows the SgrA* flare amplitude as a function 
of the flare rate 
at 2.2$\mu$m under the assumption that 
the characteristic flare duration is of the order of 100 minutes 
(see Eckart et al. 2006a). 
Longer average flare durations will shift the graph towards lower 
rates and vice versa. 
The 17$\pm$3 minute flux density 
variations lie close to the extrapolation of 
the power law line derived from flare measurements.
However, while the sub-flare variability appears to be a natural extension 
of the flare rate spectrum, there is no evidence for a large number of
2.2$\mu$m flares with durations between 100 and $\sim$20 minutes.
The sub-flare variations also lie to the right of the flare rate power spectrum
as it would be expected for any signal that is clearly discernable from the
variations imposed by the flares.
It therefore appears to be equally likely that the sub-flare variations
are due to a separate mechanism and lie beyond the possible cutoff $\kappa_2$ for
low flare amplitudes as discussed in Eckart et al. (2006a) or beyond an equally 
likely cutoff for high flare rates.
Both cutoffs can be explained within the disk model proposed by Meyer et al. (2006a)
in which the flare is due to a sound wave traveling within a finite disk. 
The disk size limitation may reflect itself in a typical flare duration and flux density
as well as a possible drop in power for shorter variations.
The higher sub-flare rates are then due to typical turbulence size scales of components within 
the disk (e.g. Hawley \& Balbus 1991, Arlt \& R\"udiger 2001).
The brightest of these orbiting components  would give rise to the observed QPOs.
The finite orbiting disk may, however, be identical to the foot point of a jet
or wind (see e.g.  Markoff, Bower \&  Falcke 2007, Markoff, Nowak \& Wilms 2005).

\subsection{A disk plus a short jet}
\label{section:diskplusjet}

\begin{figure}
   \centering
   \includegraphics[width=10cm]{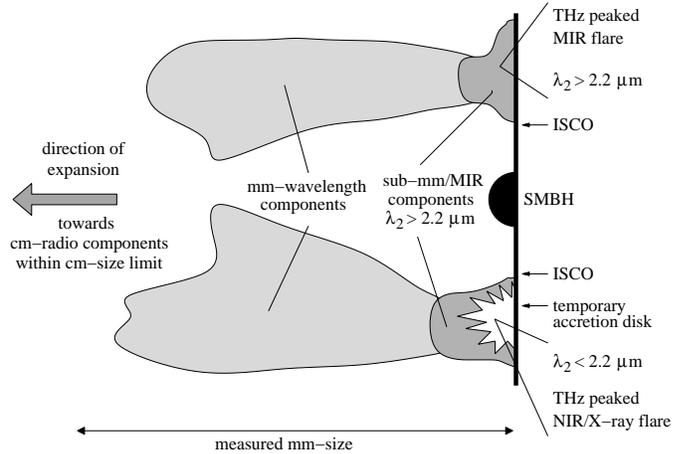}
      \caption{
%diskmodel.eps
Sketch of a possible source structure for the accretion disk around the SMBH 
associated with SgrA*.
The disk is shown as a vertical thick line to the right.
Extending to the left we show one side above the disk.
Higher energy flare emission (lower part) is responsible for the observed
NIR/X-ray flare emission.
Lower energy flare emission (upper part) may substantially contribute to long wavelength
infrared emission.
In addition to the expansion towards and beyond the the mm-source size, radial and azimuthal
expansion within the disk may occur.
Here $\lambda_2$ is the wavelength corresponding to the upper synchrotron cutoff frequency $\nu_2$.
}
         \label{fig9}
   \end{figure}

A source structure in which an accretion disk is associated with a short jet
may explain most of the observed properties of SgrA*. 
Such a configuration is sketched in Fig.\ref{fig9}.
In this figure the disk is seen edge-on. 
Details of expected jet geometries are discussed by 
Markoff, Bower \&  Falcke (2007).
We show one side above the disk and two events that 
may be characteristic for higher and lower energy flare emission. 
The higher energy events would be responsible for the observed NIR/X-ray flares.
The lower energy events would contribute most to long wavelength
infrared emission.
Higher energy events within the disk may be more stable and result in the
observed sub-flare properties.
Lower energy events may occur more detached from the disk, less stable and
more dominated by the effects of synchrotron cooling. 
While the sub-mm emission will be intimately associated with the SSC flare events
the mm-emission will originate after adiabatic expansion, further down stream the 
wind or jet emanating from the accretion disk. 
In addition radial and azimuthal
expansion of the emission zone within the disk may occur which is 
consistent with the traveling sound wave picture presented by Meyer et al. (2006a).
The sound wave or flare would be caused by the disk activity 
events depicted in Fig.\ref{fig9}.
Further expansion of the wind or jet towards more extended and diffuse
source components that dominate the cm-emission will occur.

A number of essential scenarios that comprise most of the properties 
associated with infrared/X-ray SgrA* light curves
can be explained within the model:
\\
{\bf - }I. Let us assume that the mean upper synchrotron 
cutoff lies shortward of the NIR K-band, the source 
components are stable for several cooling time scales and the
flux density variations are due to their orbital motion around SgrA*:
In this case correlated QPOs should be observed in the NIR K- and L-bands.
This situation corresponds to the orbiting spot model, 
represents events that take place within the accretion disk 
and will give rise to observed polarized infrared 
light curves that show quasi-periodic sub-flare structure.
\\
{\bf - }II. Here we assume that the mean upper synchrotron 
cutoff lies in or longward of the NIR K-band, but the 
flare events producing flux at increasingly longer wavelengths take place at increasing 
distances above the accretion plane:
In this case QPOs are preferentially observed at short NIR wavelengths.
At longer wavelengths they are less likely to occur and the 
variations are not strongly correlated with those at shorter NIR wavelengths.
This scenario would be consistent with the presence of a short jet and 
would again be valid especially for lower energy events.
\\
{\bf - }III. The mean upper synchrotron cutoff lies shortward of the NIR K-band, but the source
components are stable for only a few cooling time scales.
In this case QPOs should be observed in e.g. the NIR K- and L-bands, 
but they should be largely uncorrelated with respect to each other.
This could especially be the case for lower energy events that result in infrared 
flux density variations but are not accompanied by significant X-ray flares.
\\
{\bf - }IV. 
If the source component flux density variations are dominated by the synchrotron 
cooling time scale and are not due to relativistic effects caused by the orbital motion 
of the components around SgrA* then the K- and L-band light curves are not correlated 
with each other and no significant QPOs are observed.
Such a scenario may be observed if the flux density variations occur above the disk,
along the short jet or within the disk at larger disk 
radii.

The quasi-simultaneous K- and L-band measurements by Hornstein et al. (2007) 
make cases III and IV less likely.
Very 
weak
X-ray events have also been reported by Eckart et al. (2006a).
Here the events $\phi_1$, $\phi_2$ and $\phi_4$ could only be identified through infrared 
measurements.
In these cases the X-ray peak flux density was only of the order of 1.2 to 1.8 times the
quiescent X-ray flux density associated with Sgr~A*.
Such weak X-ray events require that the constant bremsstrahlung and variable SSC 
component of SgrA* can be distinguished through sensitive, high angular resolution X-ray 
measurements as provided by the ACIS-I instrument aboard the 
\emph{Chandra X-ray Observatory}.
However, we need to measure more flares to obtain a higher statistical 
significance.

If the entire NIR/X-ray flare event happens to 
occur above the SgrA*
accretion disk or extends a few $R_g$ into the disk corona,
then the modulation expected from an orbital spot
may be significantly reduced. Above the disk the 
spot radiation will be subject to less gravitational 
bending from the black hole and potentially 
more extended.
In that case a larger section of the underlying
disk will be heated by the X-ray flare and an increasing
amount of the lower infrared flare emission will be
inverse Compton scattered rather than synchrotron self
Compton scattered.
These effects will also lead to a significant reduction of
any QPO sub-flare contrast.

\section{Summary and Conclusion}

We have measured a significant X-ray flare that occurred synchronous to
a NIR flare with polarized sub-flares. This confirms the previous finding
(Eckart et al.  2004, 2006a, Yusef-Zadeh et al. 2006a) that there exists
a class of X-ray flares that show simultaneous NIR emission
with time lags of less than 10~minutes.
In addition there are lower energy flare events that are bright in the
infrared and are not detected in the X-ray domain (Hornstein et al. 2007).
In the relativistic disk model the May 2007 polarimetric NIR measurements 
of a flare event with the highest sub-fare contrast observed until now, 
provides direct evidence for a spot evolution during the flare.
This fact provides further strong support for the interpretation of the 
NIR polarimetry data within a relativistic disk model.
Combined with the assumption of spot expansion due to differential 
rotation the combined SSC disk model can explain the July 2004 flare
(Eckart et al. 2006a) and possibly also the 2006 July 17 flare 
reported  by Hornstein et al. (2007).

The combination of relativistic amplification curves with a simple
SSC mechanism allows us zero order interpretations in a time dependent
flare emission model.
We find that the temporary accretion disk
around SgrA* can well be represented by a multi-component model 
with source properties that are bracketed by those of a simple
flare and a quiescent model.
We have used a ($\gamma_e \sim 10^3)$ synchrotron model 
in which the source component spectral 
indices are  compatible with the constant value of $\alpha=0.6\pm0.2$ 
reported by Hornstein et al. (2007).
A steeper spectral index of $\alpha=1.3$ allows for direct synchrotron 
and SSC contributions in the NIR.
In both cases the component flux densities can be represented by
a power spectrum 
$N(S) \propto S_m^{\alpha_S}$ with an exponent $\alpha_S$ close to -1.
The multicomponent model explains the quasi-periodic
sub-flare structure at infrared wavelengths and shows that with adequate
sensitivity and time resolution they should be detectable in the X-ray domain as well.

We present a model in which a combination of a temporary accretion disk
occurs in combination with a short jet. This model can explain 
most of the properties associated with infrared/X-ray SgrA* light curves.
Simultaneous NIR K- and L-band measurements in combination with X-ray 
observations should lead to a set of light curves that should allow us
to prove the proposed model and to discriminate between the individual 
higher and lower energy flare events.
Simultaneous X-ray measurements are important to
clearly distinguish between high and low energy events.
To do so it is required to separate the thermal non-variable bremsstrahlung 
and the non-thermal variable part of the SgrA* X-ray flux density.
This capability is provided by the ACIS-I instrument aboard the 
\emph{Chandra X-ray Observatory}
and is essential to have, especially
in the case of weak X-ray flare events 
in which the X-ray flare intensity is of the 
order of the extended bremsstrahlung
component associated with SgrA* - or even below.
These can clearly be identified 
in combination with infrared data.

\begin{acknowledgements}
Part of this work was supported by the German
\emph{Deut\-sche For\-schungs\-ge\-mein\-schaft, DFG\/} via grant SFB 494.
L. Meyer, K. Muzic, M. Zamaninasab, and D. Kunneriath are members of the 
International Max Planck Research School (IMPRS) for 
Radio and Infrared Astronomy at the MPIfR and the Universities of 
Bonn and Cologne.
\end{acknowledgements}

%%%%%%%%%%%%%%%%%%%%%%%%%%%%%%%%%%%%%%%%%%%%%%%%%%%%%%%%%%%%%%

\begin{appendix}
\label{appendixA}
\noindent

\section{The multi-component disk model}
\label{multicomp}

In the following we describe an extended SSC model 
that includes a disk structure and allows us to calculate 
light curves in the NIR and X-ray domain
in order to discuss the questions posed above.

\subsection{Description of the used SSC model}
\label{section:SSCmodel}

Current models 
(Markoff et al. 2001, Yuan, Markoff \& Falcke 2002, 
Yuan, Quataert \& Narayan 2003, 2004,
Liu, Melia \& Petrosian 2006, Yuan 2006)
predict that during a flare a few percent of the electrons
near the event horizon of the central black hole are accelerated.
These models give a description of the entire electromagnetic spectrum of SgrA*
from the radio to the X-ray domain.
In contrast to these global solutions, here 
we limit our analysis to modeling the NIR to X-ray spectrum of the most compact
source component at the location of SgrA*.
Our analysis is based on a simple SSC model describing the observed 
radio to X-ray properties of SgrA* using the nomenclature given by
Gould (1979) and Marscher (1983).
Inverse Compton scattering models provide an explanation for
both the compact NIR and X-ray emission
by up-scattering sub-mm-wavelength photons into these spectral domains.
The models do not intend to explain the entire low frequency radio spectrum and 
IQ state X-ray emission.
However, they give a description of the compact emission 
from SgrA* during low and high flux density flare periods.
A more detailed explanation is also given by Eckart et al. (2004, 2006a).

We assume a synchrotron source of angular extent $\theta$. 
The source size is of the order of a few Schwarzschild
radii R$_s$=2GM/c$^2$ with 
$R_{\rm s}=1.2\times10^{10}{\rm m}$.
for a $\sim$4$\times$10$^6$\solm ~black hole. 
One R$_s$ then corresponds
to an angular diameter of $\sim$8~$\mu$as at a distance to the Galactic
Center of $\sim$8~kpc (Reid 1993, Eisenhauer et al. 2003).
The emitting source becomes optically thick at a frequency
$\nu_m$ with a flux density $S_m$, and has an optically thin spectral
index $\alpha$ following the law $S_{\nu}$$\propto$$\nu^{-\alpha}$.
The upper synchrotron cutoff frequency is $\nu_2$.
This allows us to calculate the magnetic field strength $B$ and the
inverse Compton scattered flux density $S_{SSC}$ as a function of the
X-ray photon energy $E_{keV}$ (Marscher 1983):  

$$S_{SSC} \propto ln(\nu_2/\nu_m) \theta^{-2(2\alpha+3)} \nu_m^{-(3\alpha+5)} S_m^{2(\alpha+2)} E_{KeV}^{-\alpha}~~.$$

The synchrotron self-Compton spectrum
has the same spectral index as the synchrotron spectrum that is 
up-scattered 
i.e. $S_{SSC}$$\propto$$E_{keV}$$^{-\alpha}$, and is valid within the
limits $E_{min}$ and $E_{max}$ corresponding to the wavelengths
$\lambda_{max}$ and $\lambda_{min}$ (see Marscher et al. 1983 for
further details).
We find that Lorentz factors $\gamma_e$
for the emitting electrons of the order of 
typically 10$^3$ are required to produce a sufficient SSC flux in the
observed X-ray domain.
A possible relativistic bulk motion of the emitting source results 
in a Doppler
boosting factor $\delta$=$\Gamma$$^{-1}$(1-$\beta$cos$\phi$)$^{-1}$.
Here $\phi$ is the angle of the velocity vector to the line of sight,
$\beta$ the velocity $v$ in units of the speed of light $c$, and
Lorentz factor $\Gamma$=(1-$\beta$$^2$)$^{-1/2}$ for the bulk motion.
With 'bluk motion' we mean the collective motion of the emitting 
material of an entire source component with respect to the
observer rather than the motions of the individual electrons.
In the particular case of the hot spot model it is the bulk motion 
of individual sections of the accretion disk i.e. the orbiting spot.
Relativistic bulk motion 
is not a necessity to produce sufficient SSC flux density but 
we have used modest values for 
$\Gamma$=1.2-2 and $\delta$ ranging between 1.3 and 2.0, representing a 
suitable coverage of the inclinations used in the models of the temporary
accretion disk around SgrA* (see below).
Such values will also be relevant in cases of 
mild relativistic outflows - both of which are likely to be relevant to SgrA*.

With $\gamma_e$$\sim$10$^3$, the upper synchrotron cutoff frequency 
$\nu_2$ lies within or just short-ward 
of the NIR bands such that a considerable part of the NIR spectrum 
can be explained by synchrotron emission,
and the X-ray emission by inverse Compton emission.
This is supported by SSC models presented by Markoff
et al. (2001) and Yuan, Quataert \& Narayan (2003) that result in a
significant amount of direct synchrotron emission in the infrared (see
also synchrotron models in Yuan, Quataert \& Narayan 2004 and 
discussion in Eckart et al. 2004).

\begin{table}[htb]
\begin{center}
\begin{tabular}{lcccccccr}\hline \hline
model  & $\alpha$  &$\theta$ & S$_m$ & $\nu$$_m$ \\
       &        & $\mu$as  & Jy   &  GHz     \\ \hline
$A$    & 0.80   &   3.5    &  0.40 &  1100 \\
       & 0.80   &   2.9    &  0.11 &   600 \\
       &        &          &       &       \\
$B$    & 1.30   &   6.9    &  2.40 &  1200 \\
       & 1.30   &   1.5    &  0.10 &  1100 \\
\hline \hline
\end{tabular}
\end{center}
\caption{Input parameters for the synchrotron ($A$) 
and synchrotron self Compton models ($B$).
Multi component models of the 
temporary accretion disk around SgrA* are calculated 
in \ref{section:diskmodel}
using component properties that are bracketed by 
these parameters.}
\label{minput}
\end{table}

\subsection{The SSC disk model}
\label{section:diskmodel}

In order to explain the time dependent flare properties we 
assume that the sub-flare and disk component can be described by
a number of individual synchrotron and SSC emitting source components.
Combining the light amplification curve for individual orbiting spots
and the simple SSC model described above, we can obtain zero order 
time dependent flare characteristics from the NIR to the 
X-ray domain.

As a starting point we used synchrotron models that represent a
high flux density, i.e. flaring, and a low flux density state.
Greenhough et al. (2001) outline the importance of
scaling properties of the transport processes operating within
accretion disks.
Pessah et al. (2007) present a scaling law between magnetic
stress in units of the gas pressure and
the vertical disk cell size in units of the pressure scale height
implying that the magnetic field and source component size follow a power law relation.
Therefore we assume that
the essential quantities of the SSC models, i.e. the turnover flux density
$S_m$, frequency $\nu_m$ and the source size $\theta$ 
of the individual source components are distributed as power laws
with the boundary values taken from the high and low flux density state
models. 
The exponents of the corresponding number distributions 
$N(S)$, $N(\nu)$, and $N(\theta)$ of flux components within 
the temporal accretion disk
are $\alpha_{S_m}$, $\alpha_{\nu_m}$ and $\alpha_{\theta}$:

$$N(S) \propto S_m^{\alpha_S}~~~,~~~
N(\nu) \propto \nu_m^{\alpha_\nu}~~~,~~~
N(\theta) \propto \theta^{\alpha_\theta}~~.$$

For example if $\alpha_S=0$ the flux densities of the source components cover the full
range between the minimum and maximum values. For $\alpha_S>0$ 
and $\alpha_S<0$ there is an increasing preference towards larger and lower flux 
density values, respectively. 
Similarly this is true for $\alpha_{\nu_m}$ and $\alpha_{\theta}$.

 \begin{figure}[htb]
   \centering
   \includegraphics[width=8.9cm]{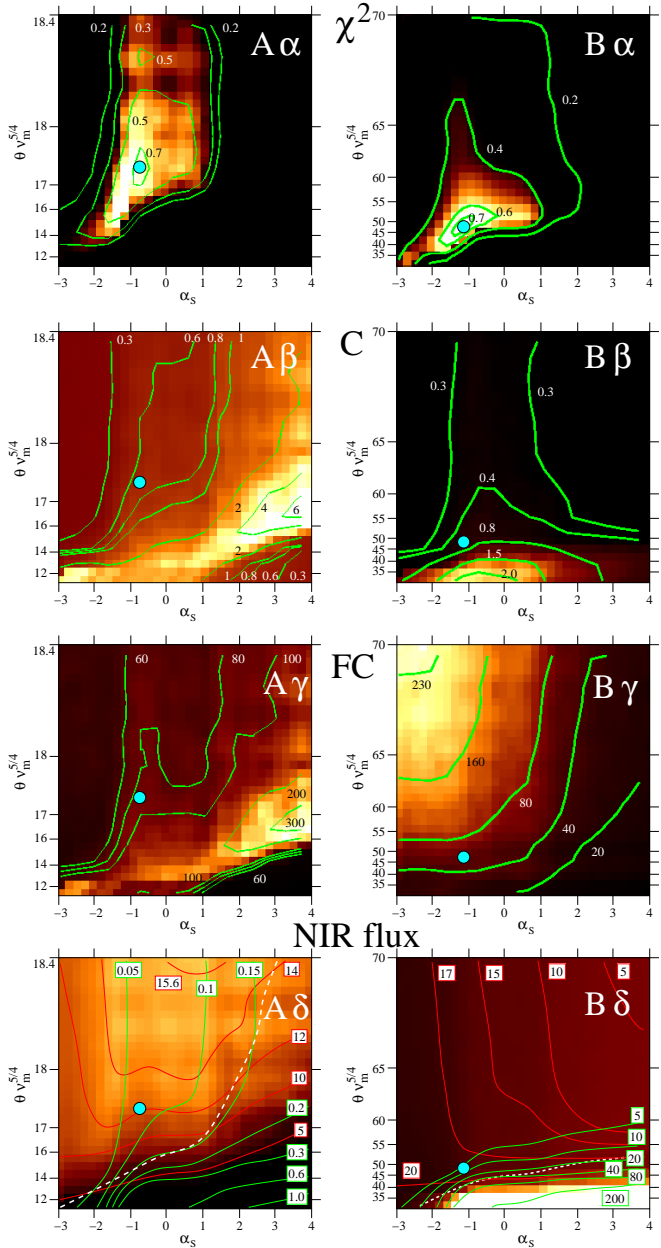}
      \caption{
%combo.eps
Diagnostic diagrams for two representative synchrotron models of 
the flare emission of SgrA*. 
}
         \label{fig4}
   \end{figure}

 \begin{figure}
   \centering
   \includegraphics[width=8cm]{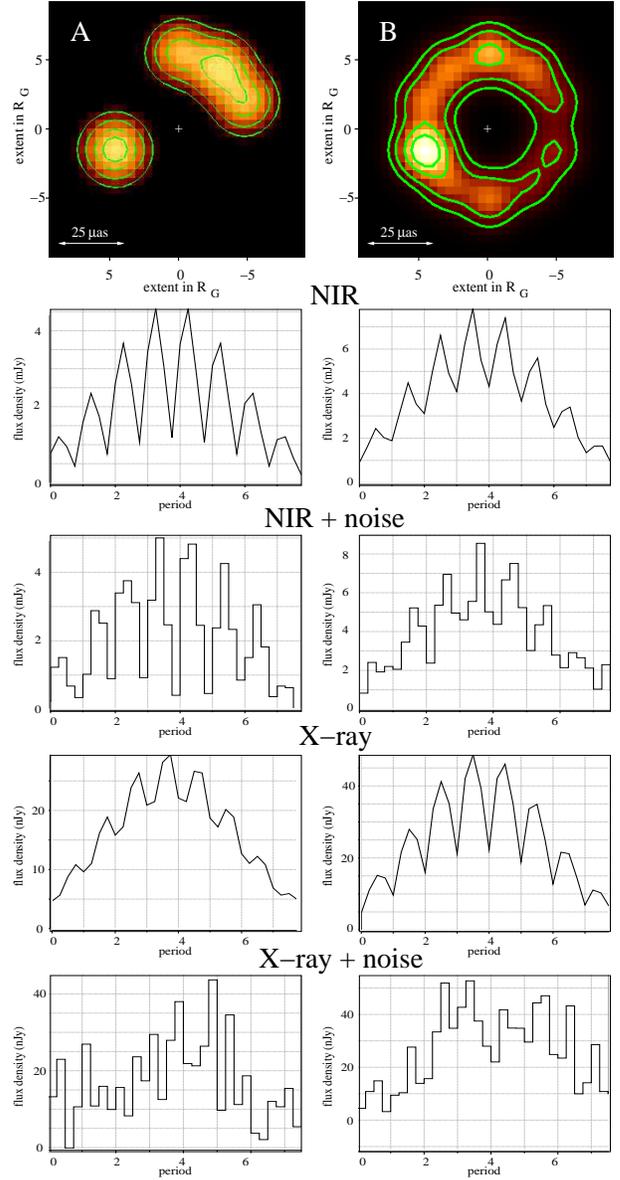}
      \caption{
%poltcurve.eps
As results from the model calculations (see Fig.\ref{fig4}) we show 
for two cases representative flux density distributions 
and NIR/X-ray light model curves with and without noise.
}
         \label{fig5}
   \end{figure}

The ISCO around a non-rotating black hole with spin parameter $a$=0 
is 6$R_g$.
% - with one Schwarzschild radius $R_s$=2$R_g$$\sim$8~$\mu$as
%with $R_s$ being the gravitational radius.
Assuming the co-rotating case that radius will shrink for higher spin parameters.
For a rotating black hole with $a$=0.5 the radius is $\sim$4.4$R_g$.
Model calculations have shown (Meyer et al. 2006a,b) 
that for SgrA* spin parameters $a\ge0.5$ and source components orbiting 
at radii larger than the ISCO are very likely. 
Meyer et al. (2006a) have shown that the disk is small with an outer disk
radius extending not much further than 2$R_s$ beyond the ISCO.
%We can therefore safely assume a radius of approximately 6~$R_g$ as a 
%characteristic size of the temporary accretion disk around SgrA*. 
With source components sizes of the order of 1.5~$R_g$ (Meyer et al. 2006a,b) 
we can therefore safely assume that the disk is well sampled 
using a total of 10 Gaussian shaped disk sections with random values of 
$S_m$, $\nu_m$ and $\theta$ taken from
the described power law distributions in order to model the entire accretion disk.
The brightest of these sections will then represent the orbiting spot
and the rest will account for the underlying disk.
This setup will of course also allow for several bright spots.
This is shown for two cases in the top panels of Fig.~\ref{fig5}.
As a simple - but still general - model 
we assumed the source components to be equally spaced 
along the circumference of a constant orbit.
While orbiting, the flux density of each component will follow the 
achromatic magnification curves that can be calculated as a function of
spin parameter $a$, inclination $i$ and orbital radius.
In addition we infer a Gaussian shaped heightening function 
with a FWHM of about 3 orbital periods, which resembles the observed 
flare lengths quite well.

In detail in Fig.~\ref{fig5} 
the flux density distributions are shown along the last stable orbit perimeter
of the super massive black hole associated with SgrA*.
Here no truncation at or just within the
last stable orbit has been applied.
We show the results for the synchrotron ($A$, left panels) and
synchrotron self Compton model ($B$, right panels).
The contour lines are at 12, 25, 50, and 75\% of the peak of the 
flux density distribution.
The NIR and X-ray light curves are representative for the median 
values at the position of the $\chi^2$ minima in Fig.4.
For comparison we added sections of the 207~s light curve in 
Fig.1 scaled to the peak values and bin size of the X-ray model light curves.
For the NIR we added 0.4~mJy of random Gaussian noise.
The bin size of the model data corresponds to 207~s for a 14~min period.
The position of SgrA* is indicated by a white cross.

As a result we obtain NIR and X-ray light curves that are modulated 
corresponding to the random distribution of component flux densities
as a function of the source component flux density $\alpha_S$ and the
product $\theta \nu_m^{5/4}$ (which is proportional to $B^{-4} S_m{^2}$).
For each pair of these quantities we calculated 100 random models and
computed median values of a number of diagnostic quantities
(see Fig.~\ref{fig4}).
Here we show the contrast of the light curve calculated as 
defined in section \ref{section:NIRObservations}.
The quantity $C$ is a measure of the flux modulation due to the presence of
sub-flares during a single flare of the characteristic duration of
about 100 minutes.
We also calculate the NIR flux density weighted magnetic field strength
times the contrast of the light curve modulation.
We take this quantity $FC$ (field contrast) as a measure of detectable 
NIR polarized flare and sub-flare structures.
Finally we show in Fig.~\ref{fig4} the NIR flux density and the
corresponding synchrotron and SSC contributions.
In detail the left and right hand panels 
in Fig.~\ref{fig4} show the
results for a synchrotron ($A$) and a synchrotron self Compton 
model ($B$), respectively.
For both models we list input parameters in Tab.~\ref{minput}.
The top two panels ($\alpha$) show the $exp(-\chi^2/2)$ results,
the middle panels ($\beta$) show the sub-flare contrast, and the
bottom panels ($\gamma$) show the NIR flux weighted magnetic field 
(see section~\label{section:SSCmodel}), and the near-infrared 
emission ($\delta$) with red and green contour lines indicating the
synchrotron and SSC contribution in that wavelength range.
The white dashed line indicates the median
short wavelength cutoff of the synchrotron spectrum. 
Significant synchrotron contribution to the total NIR flux density
occurs above  this line.
The blue filled circle in all panels indicates the location of 
the minimum $\chi^2/2$, i.e. maximum likelihood ML=$exp(-\chi^2/2)$, 
and hence the location at which the
model results shown in Fig.~\ref{fig5} have been extracted.
ML=0.5 corresponds to a 1$\sigma$ deviation.

In order to determine the agreement between the measurements and the
predicted NIR and X-ray flux density and NIR contrast C, we 
perform a maximum likelihood (ML) analysis. As a ML score 
we use 
$$log(ML) = 
- \chi^2_{S_{NIR}}/2 
- \chi^2_{S_{X-ray}}/2 
- \chi^2_{S_{C}}/2~~~.$$ 
Here  $\chi^2$ = (S$_{pred.}$-S$_{measured}$)$^2$/$\sigma$$^2$.
For the flux densities we used the values given in Tab.~\ref{flareprop}.
For the sub-flare contrast $C$ we used a value of 0.6$\pm$0.3 as derived 
from the May 2005 NIR data (Eckart et al. 2006b).
In Fig.~\ref{fig4} we show selected diagnostic diagrams for two 
representative synchrotron models of the flare emission of SgrA*. 
The models are based on the input parameters of the high and low
flux density cases listed in Tab.~\ref{minput}. These parameters 
represent the boundary values between which the component flux density, size and
peak cutoff frequency follow a power law.
We chose a spectral index of $\alpha=0.8$ for the synchrotron
model (A) in order to be compatible with the constant infrared
spectral index of $\alpha=0.6\pm0.2$ reported by Hornstein et al. (2007).
For the SSC model (B) we have chosen a spectral index of $\alpha=1.3$ 
in order to obtain a significant contribution of SSC radiation 
in the NIR bands which is impossible to achieve with flatter 
spectral indices for the given X-ray flare brightness.
The SSC X-ray flux has the same distribution as the SSC NIR flux in the 
NIR flux panel in Fig.\ref{fig4}. For the synchrotron (A in Tab.\ref{minput}) 
and SSC (B) case it is scaled down by a factor
of $\sim$1.4 and $\sim$110, respectively.

The QPO of the NIR data indicates that $a$$\ge$0.5.
We calculated the data shown in Fig.~\ref{fig4} using a spin value of
$a$=0.5 and an inclination of $i$=70$^o$ 
(Eckart et al. 2006b, see also Meyer et al. 2006ab)
and obtained $\chi^2$$>$0.9 and similar values for $i$$\ge$70$^o$.
For $a$$\rightarrow$1.0 or $i$$\rightarrow$0$^o$ the best $\chi^2$ value
drops to 0.6 and 0.3 and its location in the panels shown in Fig.\ref{fig4}
moves to the lower left.

\begin{table}
\centering
{\begin{small}
\begin{tabular}{cccccccccc}
\hline
epoch & $\Delta$$\phi$ in & radius &  $\alpha$& size    &  S$_m$ &$\nu_m$ \\
      & degree            &  $R_g$ &          &$\mu$$as$& Jy     & THz     \\
\hline
May 2007 & 0.0            &  4.0   &   0.7   &3.5   &  0.30  & 1.1  \\
         & 0.0            &  8.4   &   0.7   &3.5   &  0.40  & 1.1  \\
         &                &        &         &      &        &      \\
July 2004&  0             &  4.4   &   0.8   &3.5   &  0.40  & 1.2  \\
         &-20             &  4.6   &   0.8   &3.5   &  0.40  & 1.2  \\
         &-40             &  4.8   &   0.8   &3.5   &  0.40  & 1.2  \\
         &-60             &  6.0   &   0.8   &3.5   &  0.40  & 1.2  \\
         &-80             &  6.9   &   0.8   &3.5   &  0.40  & 1.2  \\
         &-40             &  11.3  &   0.8   &3.5   &  0.40  & 1.2  \\
         &-30             &  16.2  &   0.8   &3.5   &  0.40  & 1.2  \\
\hline
\end{tabular}
\end{small}}
\caption{Two sets of possible model parameters resulting from the multi-components calculations
for the May 2007 and July 2004 data.
We list model parameters at the beginning $t_{start}$ of the light curve sections contained in the 
red box in Fig.~\ref{simu20042007}. The exact values of the component radii and their relative
phase differences are uncertain (by about 10$^o$ in $\Delta$$\phi$ and 30\% in radius) 
as they are interdependent.
}
\label{Tab:SSCmulti}
\end{table}

\subsection{Results of the Modeling}
\label{section:modelresults}

An important result of the simulations is that the observed total NIR and X-ray flux 
densities can successfully be modeled simultaneously with the observed sub-flare contrast.
In addition the best fits to the NIR and X-ray flux densities lie within 
or close to regions of high
NIR flux density weighted magnetic field strength.
This demonstrates that the combination of the SSC modeling and the
idea of a temporary accretion disk can realistically describe the 
observed NIR polarized flares that occur synchronous with the 2-8~keV X-ray flares.
We also find that the exponential $\alpha_S$ of the assumed power law 
distribution for the synchrotron peak flux S$_m$ results in best model results for 
values around $\alpha_S$=-1$\pm$1. 
A value of $\alpha_S$=0 (which is included) 
represents scenarios in which source components cover the entire range of 
flux densities with an equal probability for each value 
rather being biased towards similarly faint or bright components.
This provides high sub-flare contrast values.
An exponent of $\alpha_S$=-1 favors lower flux density values.
In the SSC model (B) high contrast is provided by the SSC contribution 
to the NIR spectral range also allowing for $\chi^2$-fits at lower
flux density weighted magnetic field strengths around 30~G rather 
than 60~G as for the synchrotron model (A).
These magnetic fields are comparable to
the range of field strengths of the order of 0.3 Gauss to about 40 Gauss,
that we obtained in our previous model calculations 
(Eckart et al. 2004, Eckart et al. 2006ab).
The fields are also within the range expected for RIAF models
(e.g.  Markoff et al. 2001, Yuan, Quataert, Narayan 2003, 2004)
and well above the minimum value required to have the cooling time 
of the flare less than the duration of the flare 
(Yuan, Quataert, Narayan 2003, 2004, Quataert 2003).

In Fig.\ref{fig5} we show representative light curves and 
relative flux density distributions along the last stable orbit perimeter
of the super massive black hole associated with SgrA*
for the synchrotron (A) and SSC (B) model.

While a single spot model, including a disk contribution,
is successful in explaining the observed infrared polarized light curves, 
a model with two spots located at opposite sites of the temporary accretion 
disk may appear as an attractive solution as well.
Such models are motivated by the possible formation of spiral arms within such a disk 
that may result in corresponding 'hot spots' at the position at which the spiral arms 
originate and are closest to the last stable orbit.
Modulation now occurs with the pattern speed rather than the orbital
velocity.
This is the case for the Rossby wave instability as discussed by 
Falanga et al. (2007).

However, a comparison between the two panels shows that for spot sizes 
of the order of one $R_g$ and above such a two spot model is not very well applicable.  
Due to these spot sizes and due to the fact that (especially for extended spots) the boosting is a 
rather slow function of the spots position on its orbit, the corresponding 
amplification curves of an individual spot 
have full widths at half maxima that are in the range of 
0.3 to 0.5 of a single orbital period.
Such shallow amplification curves result in a rather low modulation contrast
of $<$0.2 for spots of equal brightness, especially if they are located on opposite sides.
If they have different brightnesses the contrast changes in favor of a 
single spot model as discussed in previous papers 
(Eckart et al. 2006b, Meyer et al. 2006a,b, Broderick \& Loeb 2006a,b).
Similarly a model of equally bright spots orbiting at radii larger than the
last stable orbit will result in a low modulation contrast, since the 
amplification is a function of the orbital velocity ($\propto$$r^{-0.5}$)
and therfore suffers from an 
additional decrease ($\Gamma(r)$$\rightarrow$1) in boosting.
In summary: Among orbital spot models with spot sizes of the order of $\sim$1.5~$R_g$,
the observed light curve modulations with a contrast of 0.3 to 0.6 and above,
favor scenarios with a single dominant bright spot.
To demonstrate that it is feasible to give a first order description of the 
NIR and X-ray light curves with the multi-component approach, 
we applied the model calculations to the 
July 2004 data (Eckart et al. 2006a)
July 2005 and May 2007 (presented here).
The calculations have been performed for a spin parameter of $a$=0.5 
(as a save lower limit to the spin parametrer;
see Genzel et al. 2003, Eckart et al. 2006b, Meyer et al. 2006ab, 2007)
and a disk inclination of $i$=70$^o$.
The results are summarized in section~\ref{multicomp} and in 
Tab.~\ref{Tab:SSCmulti}

\end{appendix}

\end{document}